\documentclass[twocolumn,10pt]{article}

\usepackage{graphicx,amsmath,amssymb,url,subfigure,bibspacing,tweaklist,usenix}

\renewcommand{\leq}{\leqslant}

\long\def\comment#1{}

\begin{document}

\title{
    Lessons from Three Views of the Internet Topology
}

 \author{
    Priya Mahadevan \\ UCSD/CAIDA \\ pmahadevan@cs.ucsd.edu  \and
    Dmitri Krioukov \\ CAIDA \\ dima@caida.org \and
    Marina Fomenkov \\ CAIDA \\ marina@caida.org \and
    Bradley Huffaker \\ CAIDA \\ brad@caida.org \and
    Xenofontas Dimitropoulos \\ Georgia Tech \\ fontas@caida.org \and
    kc claffy \\ CAIDA \\ kc@caida.org \and
    Amin Vahdat \\ UCSD \\ vahdat@cs.ucsd.edu 
}

\date{}

\maketitle

\begin{abstract}

\noindent Network topology plays a vital role in understanding the
performance of network applications and protocols.  Thus, recently
there has been tremendous interest in generating {\em realistic}
network topologies.  Such work must begin with an understanding of
existing network topologies, which today typically consists of a
relatively small number of data sources.  In this paper, we calculate
an extensive set of important characteristics of Internet AS-level
topologies extracted from the three data sources most frequently used
by the research community: traceroutes, BGP, and WHOIS.  We find that
traceroute and BGP topologies are similar to one another but differ
substantially from the WHOIS topology.  We discuss the interplay
between the properties of the data sources that result from specific
data collection mechanisms and the resulting topology views.  We find
that, among metrics widely considered, the {\em joint degree
distribution} appears to fundamentally characterize Internet
AS-topologies: it narrowly defines values for other important
metrics. We also introduce an evaluation criteria for the accuracy of
topology generators and verify previous observations that generators
solely reproducing degree distributions cannot capture the full
spectrum of critical topological characteristics of any of the three
topologies. Finally, we release to the community the input topology
datasets, along with the scripts and output of our calculations. This
supplement should enable researchers to validate their models against
real data and to make more informed selection of topology data sources
for their specific needs.

\end{abstract}

\section{Introduction}\label{sec:intro}

Internet topology analysis and modeling has attracted substantial
attention recently~\cite{FaFaFa99,WilRevisited,
TaGoJaShWi02,LiAlWiDo04,BuTo02,JaRoTo04,GaPa04,ZhoMo04}.\footnote{We
intentionally avoid citing statistical physics literature, where
the number of publications dedicated to the subject has exploded.
For introduction and references
see~\cite{DorMen-book03,nets-nr-won}.} Such an interest is not
surprising since the Internet's topological properties and their
evolution are cornerstones of many practical and theoretical
network research agendas. Our own motivation for this study is the
need to construct accurate network emulation
environments~\cite{vahdat02} that will enable development,
reliable testing, and performance evaluation of new applications,
protocols, and routing architectures~\cite{nets-nr-won}. Knowledge
of realistic network topologies and the availability of tools to
generate them are essential to this goal.
We also seek to develop a methodology to
compare topologies to one another based on relatively simple metrics.
That is, we seek a set of metrics such that when two topologies
demonstrate similar values for a particular property, they will be
similar across a broad range of potential properties.

There are a number of sources of Internet topology data, obtained
using different methodologies that yield substantially different
topological views of the Internet. Unfortunately, many researchers either rely
only on one data source, sometimes outdated or incomplete, or mix disparate
data sources into one topology. To date, there has been little attempt to provide a
detailed analytical comparison of the most important properties of topologies
extracted from the different data sources.

Our study fills this gap by analyzing and explaining topological properties
of Internet AS-level graphs extracted from the three commonly used data sources:
(1)~traceroute measurements~\cite{skitter}; (2)~BGP~\cite{routeviews}; and (3)~the WHOIS database~\cite{irr}.
This work makes three key contributions to the field of topology research:
\begin{enumerate}
\item We calculate a broad range of topology metrics considered in the
networking literature for the three sources of data. We reveal the peculiarities of each data source and the
resulting interplay between artifacts of data collection and the key properties
of the derived graphs.
\item We highlight the interdependencies
between a broad array of topological features and discuss their
relevance when comparing Internet topologies to various random graph
models that attempt to capture Internet topology characteristics.  Our
analysis shows that graph models that reproduce the joint degree
distribution of the graphs also capture other crucial topological
characteristics to best approximate the topology.

\item To promote and simplify further analysis and discussion, we
release~\cite{comp-anal} the following data and results to the community:
a)~the AS-graphs representing the topologies extracted from the raw data sources;
b)~the full set of data plots (many not included in the paper) calculated for
all graphs;
c)~the data files associated with the plots, useful for researchers
looking for other summary statistics or for direct comparisons with
empirical data;
and d)~the scripts and programs we developed for our calculations.
\end{enumerate}

We organize this paper as follows. Section~\ref{sec:graphs} describes our data
sources and how we constructed AS-level graphs from these data.
In Section~\ref{sec:characteristics} we present the set of topological
characteristics calculated from our graphs and explain what they measure and
why they are important. Section~\ref{sec:modelcomp} compares properties of the
observed topologies
with classes of random graphs and discusses the
accuracy criteria for topology generators. We discuss the limitations of our study in Section~\ref{sec:limitations}. We conclude in
Section~\ref{sec:conclusion} with the summary of our findings.

\section{Construction of AS-level graphs}
\label{sec:graphs}

\subsection{Data sources}
\label{sec:data_sources}

We used the following
data sources to construct AS-level graphs of the Internet: traceroute measurements, BGP data, and the WHOIS
database.

{\bf Traceroute}~\cite{traceroute} is a tool that captures a sequence of IP
hops along the forward path from the source to a given destination by sending
either UDP or ICMP probe packets to the destination.

CAIDA has developed a
tool, {\em skitter}~\cite{skitter}, to collect continuous traceroute-based Internet
topology measurements. AS-level topology graphs derived from the {\em skitter}
data on a daily basis are available for download at~\cite{as-adjacencies}.
For this study, we used the 31 daily graphs for the month of
March~2004.
The measurements contain multi-origin ASes
(prefixes announced by different originating ASes)~\cite{MaReWaKa03},
AS-sets~\cite{bgp}, and private ASes~\cite{as-guidelines}.
Both multi-origin ASes and AS-sets create ambiguous mapping between
IP addresses and ASes, while private ASes create false links.
Hence we filter AS-sets, multi-origin ASes,
and private ASes from each graph, and we discard indirect
links~\cite{as-adjacencies}. We then merge the each daily graph to form one graph
referred to as the {\em skitter graph} throughout the rest of the paper.

{\bf BGP} (Border Gateway Protocol)~\cite{bgp} is the protocol used for routing
among ASes in the Internet.
RouteViews~\cite{routeviews} collects and archives both static snapshots of the
BGP routing tables and dynamic BGP data in the form of BGP message dumps
(updates and withdrawals). Therefore, we derive two types of
graphs from the BGP data for the same month of March~2004: one from the static
tables ({\bf BGP tables}) and one from the updates ({\bf BGP updates}). In both
cases, we filter AS-sets and private ASes and merge the 31 daily graphs
into one.

{\bf WHOIS}~\cite{irr} is a collection of databases containing a wide range of
information useful to network operators. Unfortunately, these databases are
manually maintained with little requirements for updating the registered
information in a timely fashion. RIPE's~\cite{SiFa04,ChaGoJaSheWi04} WHOIS database contains the most reliable current topological
information, although it covers primarily European Internet infrastructure.

We obtained the RIPE WHOIS database dump for April~07, 2004. The records of
interest to us are: {\tt
\begin{center}
\begin{tabular}{ll}
aut-num:&ASx\\
import:&from ASy\\
export:&to ASz
\end{tabular}
\end{center}}
\noindent which indicate links {\tt ASx-ASy} and {\tt ASx-ASz}. We construct
an AS-level graph (referred to as {\em WHOIS graph}) from these data and exclude
ASes that did not appear in the {\tt aut-num} lines. Such ASes are external to
the database and we cannot correctly estimate their topological properties
(e.g.~node degree). We also filter private ASes.

All four graphs constructed as described are available for download
from~\cite{comp-anal}. Overlap statistics of the graphs are shown in
Table~\ref{table:comparison}.

\begin{table*}[tbh]
    \centering
    \caption{{\bf Comparison of graphs built from different data sources.}
    The baseline graph~\mbox{$G_A$} is the BGP-tables graph.
    Graph~\mbox{$G_B$} is one the other graphs listed in the first row.}
    \begin{tabular}{|l|c|c|c|}
\hline
&  BGP updates&  skitter&  WHOIS\\ \hline
 Number of nodes in both $G_A$ and $G_B$   ($|V_{A} \bigcap V_{B}|$) & 17,349 & 9,203 & 5,583 \\  
\hline
 Number of nodes in $G_A$ but not in $G_B$ ($|V_{A} \setminus V_{B}|$) & 97 & 8,243 & 11,863 \\  
\hline
 Number of nodes in $G_B$ but not in $G_A$ ($|V_{B} \setminus V_{A}|$) & 68 & 1 & 1,902 \\  
\hline
 Number of edges in both $G_A$ and $G_B$   ($|E_{A} \bigcap E_{B}|$) & 38,543 & 17,407 & 12,335 \\  
\hline
 Number of edges in $G_A$ but not in $G_B$ ($|E_{A} \setminus E_{B}|$) & 2,262 & 23,398 & 28,470 \\  
\hline
 Number of edges in $G_B$ but not in $G_A$ ($|E_{B} \setminus E_{A}|$) & 3,941 & 11,552 & 44,614 \\  
\hline
\end{tabular}

    \label{table:comparison}
\end{table*}

Comparing the two BGP-derived graphs, we note that the sets of their
constituent nodes and links are similar. Given minor differences between node and link sets of the BGP table- and
update-derived topologies, we, not surprisingly, found the metric values calculated for these two graphs to be close. Therefore, in the rest of this study we present characteristics of the
static BGP-table graph only and refer to it as {\em BGP graph}.\footnote{Plots
and tables with metrics of the BGP-update graph included are available in the
Supplement~\cite{comp-anal}.}

In constructing the skitter graph, we used BGP tables to map IP addresses
observed in traceroutes to AS numbers. Therefore the number of nodes seen by
skitter but not by BGP should be~$0$. The one node difference (AS2277 Ecuanet
in skitter data) results from the fact that different BGP table dumps were used to construct the BGP-table graph and to map an IP address to this AS on the day when skitter observed this IP address in its traces.

Based on the very method of their construction, the three graphs in
this study reveal different sides of the actual Internet AS-level
topology. The skitter graph closely reflects the topology of actual
Internet traffic flows, i.e.~the data plane. The BGP graph reveals the
topology seen by the routing system, i.e.~the control plane.  However,
both skitter and BGP are {\em traceroute-like\/} explorations of the
network topology, meaning that we can try to approximate these graphs
by a union of spanning trees rooted at, respectively, skitter monitors
or BGP data collection points. As such, both these methods discover
more {\bf radial} links, that is, links connecting numerous low-degree
nodes (e.g.~customers ASes) to high-degree nodes (e.g.~large ISP
ASes). At the same time, these measurements fail to detect many {\bf
tangential}\footnote{The semantics behind the terms ``radial'' and
``tangential'' come from the skitter poster layout~\cite{skitter-poster}, where high-degree nodes populate the
center of a circle, while low degree nodes are close to the
circumference. Links connecting high-degree nodes to low-degree nodes
are indeed radial then.} links, that is, links between nodes of
similar degrees.  Traceroute-like methods are particularly unsuitable
for discovering tangential links interconnecting medium-to-low degree
nodes (e.g.~lower-tier ASes) since many of these links do not lie on
any shortest path rooted at a particular vantage point in the core.
In contrast, WHOIS data contains abundant medium-degree tangential
links as directly attached to sources of WHOIS records (values of {\tt
aut-num} fields).

\subsection{Statistical validity of our results}\label{sec:stat_val}

Lakhina {\em et al.}~\cite{LaByCroXie03} numerically explored sampling biases
arising from traceroute measurements and found that such traceroute-sampled
graphs of the Internet yield insufficient evidence for characterizing the
actual underlying Internet topology. However, Dall'Asta
{\em et al.}~\cite{DaAlHaBaVaVe05} convincingly refute their conclusions
by showing that various traceroute exploration strategies provide sampled
distributions with enough signatures to distinguish at the statistical level
between different topologies. The authors of~\cite{DaAlHaBaVaVe05} also argue
that real mapping experiments
observe genuine features of the Internet, rather than artifacts.
These results lend credibility to our chosen traceroute-like data
sources and imply that the real Internet topology is unlikely to be
critically different from the ones measured in skitter and BGP cases.

The topology metrics we consider in Section~\ref{sec:characteristics} all show
that the WHOIS topology is different from the other two graphs. Thus,
the following
question arises: Can we explain the difference by the fact that the WHOIS graph
contains only a part of the Internet, namely European ASes? To answer this
question we performed the following experiment. We considered the BGP-tables
and WHOIS topologies narrowed to the set of nodes present both in BGP tables
and WHOIS (cf.~Table~\ref{table:comparison}) and compared the various
topological characteristics for the full and the reduced graphs. Results of
this comparison are available in the Supplement~\cite{comp-anal}. We found
that the induced graphs preserve the full set of the properly normalized
topological properties of the original graphs. Therefore, the differences
between full BGP and WHOIS topologies are intrinsic to their originating data
sources, and not due to geographical biases in sampling the Internet.

\section{Topology characteristics}\label{sec:characteristics}

In this section, we quantitatively analyze differences between
the three graphs in terms of various topology metrics. The set of metrics we
discuss here encompasses most of the graph metrics considered relevant
for topology  in the networking literature~\cite{TaGoJaShWi02,LiAlWiDo04,BuTo02,ZhoMo04}.
Relative to most related work, we consider a broader array of metrics
of interest.

For each metric, we address the following points: 1)~metric definition;
2)~metric importance; and 3)~discussion on the metric values for the three measured topologies. We present these results in the plots associated
with every metric and in the master Table~\ref{table:summary} containing all the scalar metric values for all the three graphs.

We start with simple and basic metrics that characterize local connectivity in a network. With increasing precision, we move on to more sophisticated
metrics that describe global properties of the topology. The latter metrics play a vital role in the performance of network protocols and applications.
Some metrics that we
discuss here are not exactly equal but directly related to
a topology characteristic deemed important in the networking literature.
Where possible, we illuminate the relationship between the metrics we consider and the ones that have been discussed in influential networking papers.
We provide a summary of this mapping in Table~\ref{table:importance}.

\begin{table}[h]
    \centering
    \caption{\bf Important metric mappings.}
    \begin{tabular}{|l|l|}
      \hline
      Previously defined metric& Our definition \\
      \hline\hline
      {\em Likelihood\/} in~\cite{LiAlWiDo04} & Assortativity coefficient \\
      \hline
      {\em Expansion\/} in~\cite{TaGoJaShWi02} & Distance \\
      \hline
      {\em Resilience\/} in~\cite{TaGoJaShWi02} & \\
      {\em Performance\/} in~\cite{LiAlWiDo04} & \raisebox{1.4ex}[0pt]{Spectrum} \\
      \hline
      {\em Link value\/} in~\cite{TaGoJaShWi02} & \\
      {\em Router utilization\/} in~\cite{LiAlWiDo04} & \raisebox{1.4ex}[0pt]{Betweenness} \\
      \hline
    \end{tabular}
    \label{table:importance}
\end{table}

\subsection{Average degree}

\textbf{\textit{Definition.}}
The two most basic graph properties are the
{\bf number of nodes}~$n$ (also referred as {\bf graph size}) and the
{\bf number of links}~$m$. They define
the {\bf average node
degree}~\mbox{$\bar{k}=2m/n$}.

\textbf{\textit{Importance.}}
Average degree is the coarsest connectivity
characteristic of the topology. Networks with higher~$\bar{k}$ are ``better-connected'' on average
and, consequently, are likely to be more robust.
Detailed topology characterization based only on the average degree is rather limited, since graphs with the same average node degree can have vastly different structures.

\textbf{\textit{Discussion.}}
BGP sees almost twice as many nodes as skitter
(Table~\ref{table:summary}).
The WHOIS graph is smallest,
but its average degree is almost three times larger than that of BGP, and
\mbox{$\sim 2.5$} times larger than that of skitter. In other words, WHOIS contains
substantially more links, both in the absolute~($m$) and relative~($\bar{k}$)
senses, than any other data source, but credibility of these links is
lowest (cf.~Section~\ref{sec:data_sources}): there have been reports about some ISPs that tend to enter inaccurate information in the WHOIS database in order increase their ``importance'' in the Internet hierarchy~\cite{SiFa04}.

Graphs ordered by increasing average degree~$\bar{k}$ are BGP,
skitter, WHOIS. We call this order the {\bf $\mathbf{\bar{k}}$-order}.

\subsection{Degree distribution}\label{sec:degree-distr}

\textbf{\textit{Definition.}}
Let~$n(k)$ be the number of nodes of degree~$k$
($k$-degree nodes). The {\bf node degree
distribution} is the probability that a randomly selected
node is $k$-degree: \mbox{$P(k)=n(k)/n$}. The degree distribution
contains more information about
connectivity in a given graph than the average degree, since given
a specific form of~$P(k)$ we can always restore the average degree by
\mbox{$\bar{k} = \sum_{k=1}^{k_{max}} kP(k)$},
where~$k_{max}$ is the {\bf maximum node degree} in the graph. If the
degree distribution in a graph of size~$n$ is a power law, \mbox{$P(k) \sim k^{-\gamma}$},
where~$\gamma$ is a positive {\bf exponent}, then~$P(k)$ has a natural
cut-off at the {\bf power-law maximum degree}~\cite{DorMen-book03}:
\mbox{$k_{max}^{PL} = n^{1/(\gamma-1)}$}.

\textbf{\textit{Importance.}}
The degree distribution is the most frequently used topology characteristic.
The observation~\cite{FaFaFa99} that the Internet's degree distribution follows
power law had significant impact on network topology research:
Internet models before~\cite{FaFaFa99} failed to exhibit power
laws. Since power law is a highly variable distribution, node degree
is an important attribute of an individual node. For example, we can
use AS degrees as the simplest way to rank ASes~\cite{as-ranking}.

\begin{figure*}[tbh]
    \centerline{
        \subfigure[PDF]
        {\includegraphics[width=2.2in]{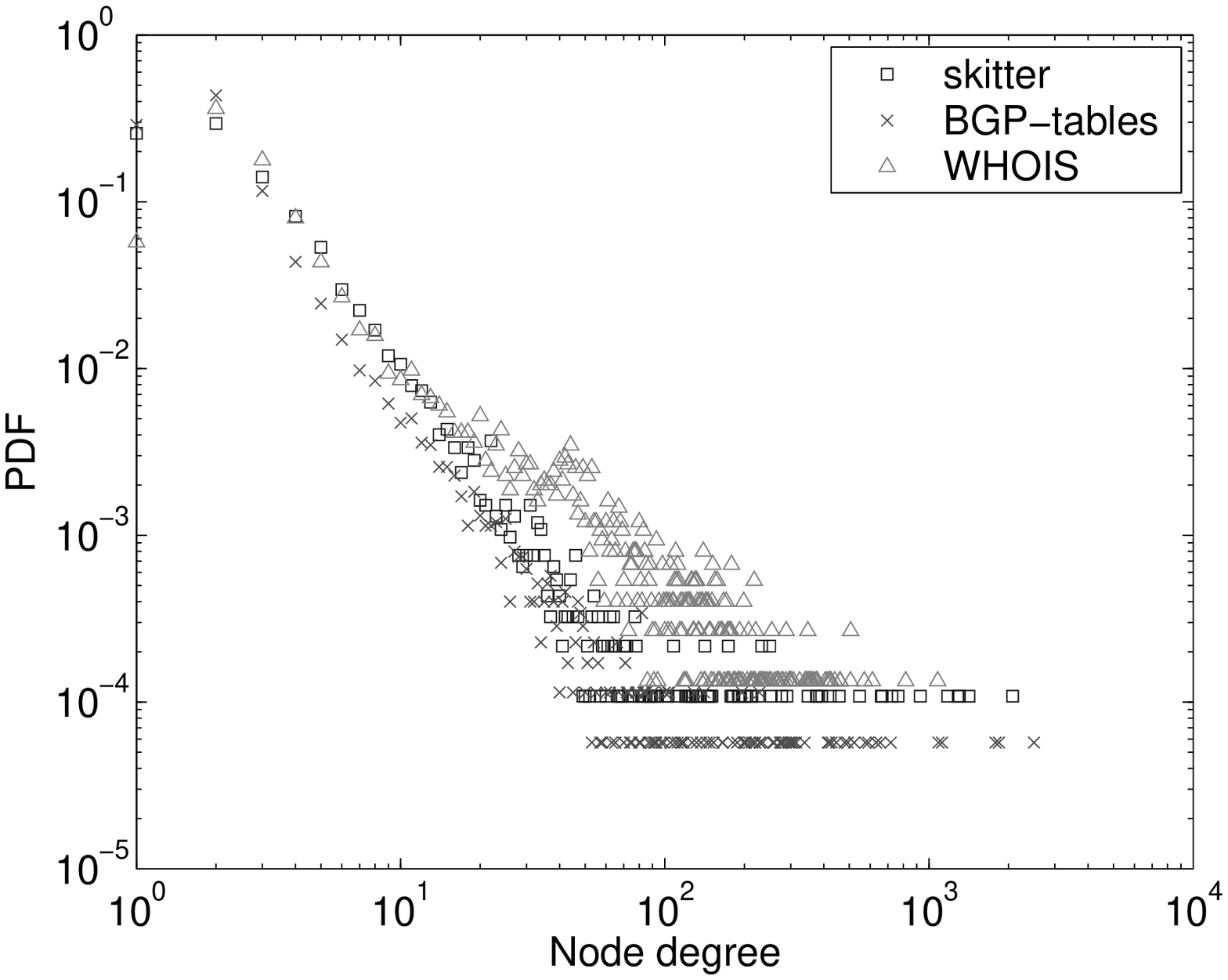}
        \label{fig:pk-pdf}}
        \hfill
        \subfigure[CCDF]
        {\includegraphics[width=2.2in]{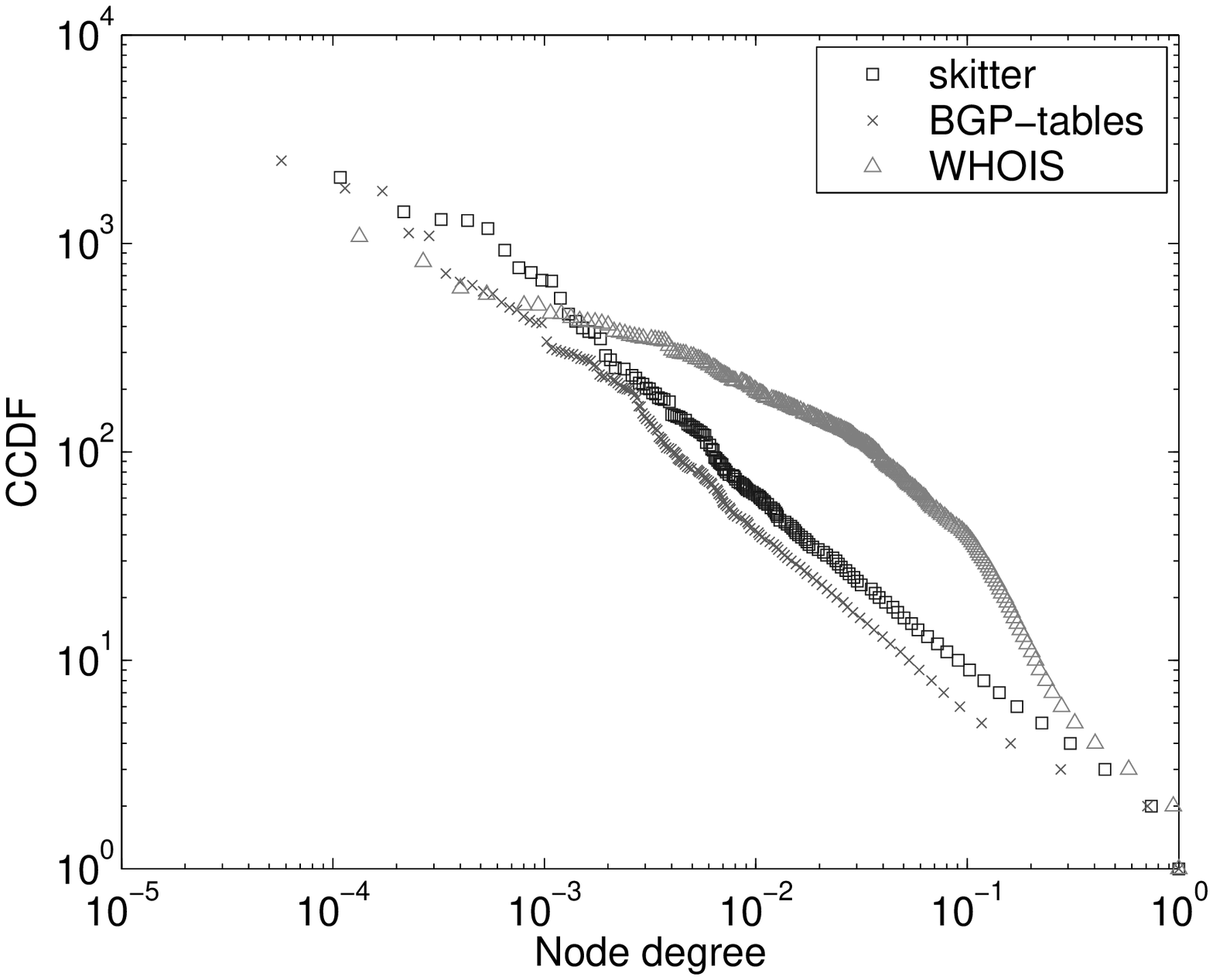}
        \label{fig:pk-ccdf}}
    \subfigure[PDFs of skitter vs.\ BGP differences]
        {\includegraphics[width=2.5in]{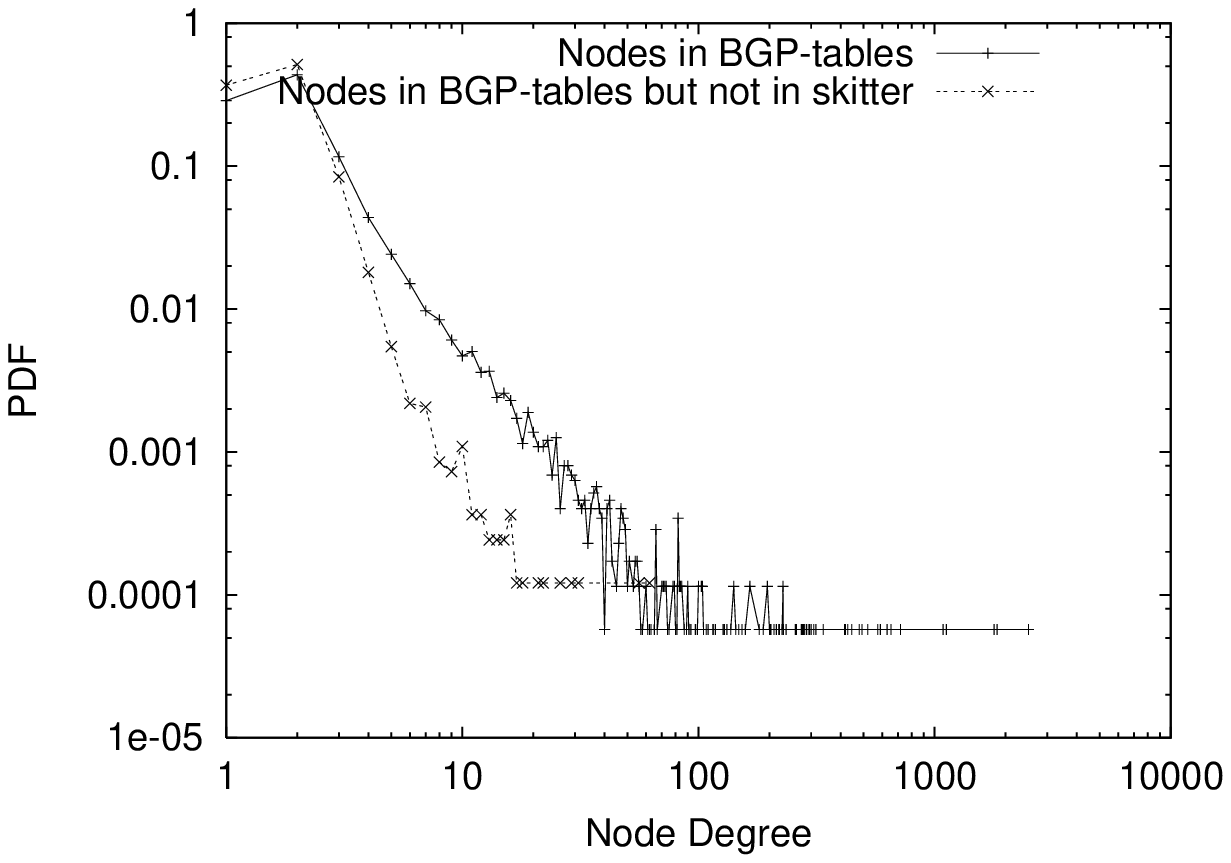}
        \label{fig:sk_vs_bgp:pk}}
    }
    \caption{\footnotesize \bf Node degree distributions~$\mathbf{P(k)}$.
}
    \label{fig:pk}
\end{figure*}

\textbf{\textit{Discussion.}}
As expected, the degree distribution PDFs and CCDFs
in Figure~\ref{fig:pk} are in the
$\bar{k}$-order (BGP $<$ skitter $<$ WHOIS) for a wide range of
node degrees.

Comparing the observed maximum node degrees~$k_{max}$ with those predicted
by the power law~$k_{max}^{PL}$ in Table~\ref{table:summary}, we conclude
that skitter is closest to power law. The power-law approximation for the BGP
graph is less accurate. The WHOIS graph has an excess of medium degree nodes
and its node degree distribution does not follow power law at all. It is not
surprising then that augmenting the BGP graph with WHOIS links breaks the power
law characteristics of the BGP graph~\cite{WilRevisited,ChaGoJaSheWi04}.

Note that there are fewer 1-degree nodes than 2-degree nodes in all the graphs
(cf.~Figure~\ref{fig:pk-pdf}). This effect is due to
the AS number assignment policies~\cite{as-guidelines}
allowing a customer to have an AS number only if it has multiple providers.
If these policies were strictly enforced, then the minimum AS degree would be~2.

CCDFs of skitter and BGP graphs look rather similar (Figure~\ref{fig:pk-ccdf}),
but Table~\ref{table:comparison} shows significant differences between the two
graphs, in terms of (non-)intersecting nodes and links. We seek to answer the
question of where, topologically, these nodes and links are located.
Calculating the degree distribution of nodes present only in the BGP
graph (Figure~\ref{fig:sk_vs_bgp:pk}), we detect a skew towards low-degree nodes.
The average degree of the nodes that are present only in BGP graphs, and not in skitter is~$1.86$. Skitter's target list of
destinations to probe does not contain any replying IP address in the address
blocks advertised by these small ASes. As a result, the skitter graph misses
them.

Most links present only in BGP, but not in skitter, are tangential links between
low-degree ASes (see~\cite{comp-anal} for details). The majority of such links
connect the low-degree ASes present only in BGP to their secondary (backup)
low-degree providers, while their primary providers are of high degrees.
Even if skitter detects
a low-degree AS having such a small backup provider, this tool is still unlikely
to detect the backup link since its traceroutes follow the primary path via the
large provider.

\subsection{Joint degree distribution}\label{sec:joint-degree}

While the node degree distribution tells us how many nodes of given degree are in
 the network, it fails to provide information on the interconnection between these nodes: given~$P(k)$, we still do not know anything about
the structure of the neighborhood of the average node of a given degree.
The joint degree distribution (or degree-degree correlation matrix) fills this gap by providing information about nodes' 1-hop neighborhoods.

\textbf{\textit{Definition.}}
Let~$m(k_1,k_2)$ be the total number of edges connecting nodes of
degrees~$k_1$ and~$k_2$. The {\bf joint degree
distribution} (JDD) is the probability that a randomly selected
edge connects $k_1$- and $k_2$-degree nodes:
\mbox{$P(k_1,k_2) \sim m(k_1,k_2)/m$}.\footnote{The exact definition for
undirected graphs differentiates (by a factor~$1/2$) between the
\mbox{$k_1 = k_2$} and \mbox{$k_1 \neq k_2$} cases.}
Note that $P(k_1,k_2)$ is different from the conditional
probability~$P(k_2|k_1)=\bar{k}/k_1P(k_1,k_2)/P(k_1)$ that a
given $k_1$-degree node is connected to a $k_2$-degree node.
The JDD contains more information about
the connectivity in a graph than the degree distribution, since given a
specific form of~$P(k_1,k_2)$ we can always restore both the degree
distribution~$P(k)$ and average degree~$\bar{k}$ by expressions
in~\cite{DorMen-book03}. The JDD is a function of two arguments. A summary
statistic of JDD, that is a function of one argument is called
{\bf the average neighbor
connectivity}~\mbox{$k_{nn}(k) = \sum_{k'=1}^{k_{max}} k' P(k'|k)$}.
It is simply the average neighbor degree of the average $k$-degree node.
It shows whether ASes of a given degree preferentially
connect to high- or low-degree ASes.
In a full mesh graph, $k_{nn}(k)$ reaches its maximal possible value $n-1$.
Therefore, for uniform graph comparison we plot normalized values $k_{nn}(k)/(n-1)$.
We can further summarize JDD by a single scalar called
{\bf assortativity coefficient}~$r$~\cite{newman02,dorogovtsev03}.

\textbf{\textit{Importance.}}
As opposed to the degree distribution, the network community has recently started recognizing the importance on JDD~\cite{WiJa02,JaRoTo04}.
The most prominent recent example defines {\em likelihood}~\cite{LiAlWiDo04}---the central metric for their argument---as
a metric directly related to the assortativity coefficient. They propose to use likelihood as a measure of randomness
differentiating between multiple
graphs with the same degree distribution. Such a measure is important
for evaluating the amount of order (e.g.~engineering design constraints) present in a given topology. A topology with low likelihood is not random,
it is a result of some sophisticated evolution processes involving specific
design purposes. We actively use the JDD in the described fashion in
Section~\ref{sec:modelcomp}.

The assortativity coefficient $r$ ( \mbox{$-1 \leq r \leq 1$}) has direct practical implications. {\bf Disassortative}
networks with \mbox{$r<0$} have an excess radial links connecting nodes of
dissimilar degrees. Such networks are vulnerable to both random
failures and targeted attacks. Viruses spread faster in these topologies. On a  positive side, vertex cover in disassortative graphs is smaller, which is important for applications such as traffic monitoring~\cite{BreChaGaRaSi01} and prevention of DoS attack~\cite{PaLe01}. The opposite properties apply to {\bf assortative}
networks with \mbox{$r>0$} that have an excess of tangential links connecting nodes of
similar degrees.

\textbf{\textit{Discussion.}}
All the three Internet
graphs built from our data sources are disassortative (\mbox{$r<0$}) as seen in
Table~\ref{table:summary}. We call the order of graphs with decreasing assortativity
coefficient~$r$---WHOIS, BGP, skitter---the {\bf $\mathbf{r}$-order}.
The most disassortative graph is skitter, that has the largest
excess of radial links. The least disassortative graph is WHOIS. The $r$-order can be explained in terms of differing topology measurement methodologies. As described in Section~\ref{sec:graphs}, the traceroute-like explorations of BGP and skitter data fail to detect tangential links, thus causing the graphs to be disassortative. The WHOIS graph's collection methodology however finds abundant medium-degree tangential links, resulting in the graph's higher assortative value.

\begin{figure*}[tbh]
  \begin{minipage}[t]{2.2in}
      \centerline{
          \includegraphics[width=2.2in]{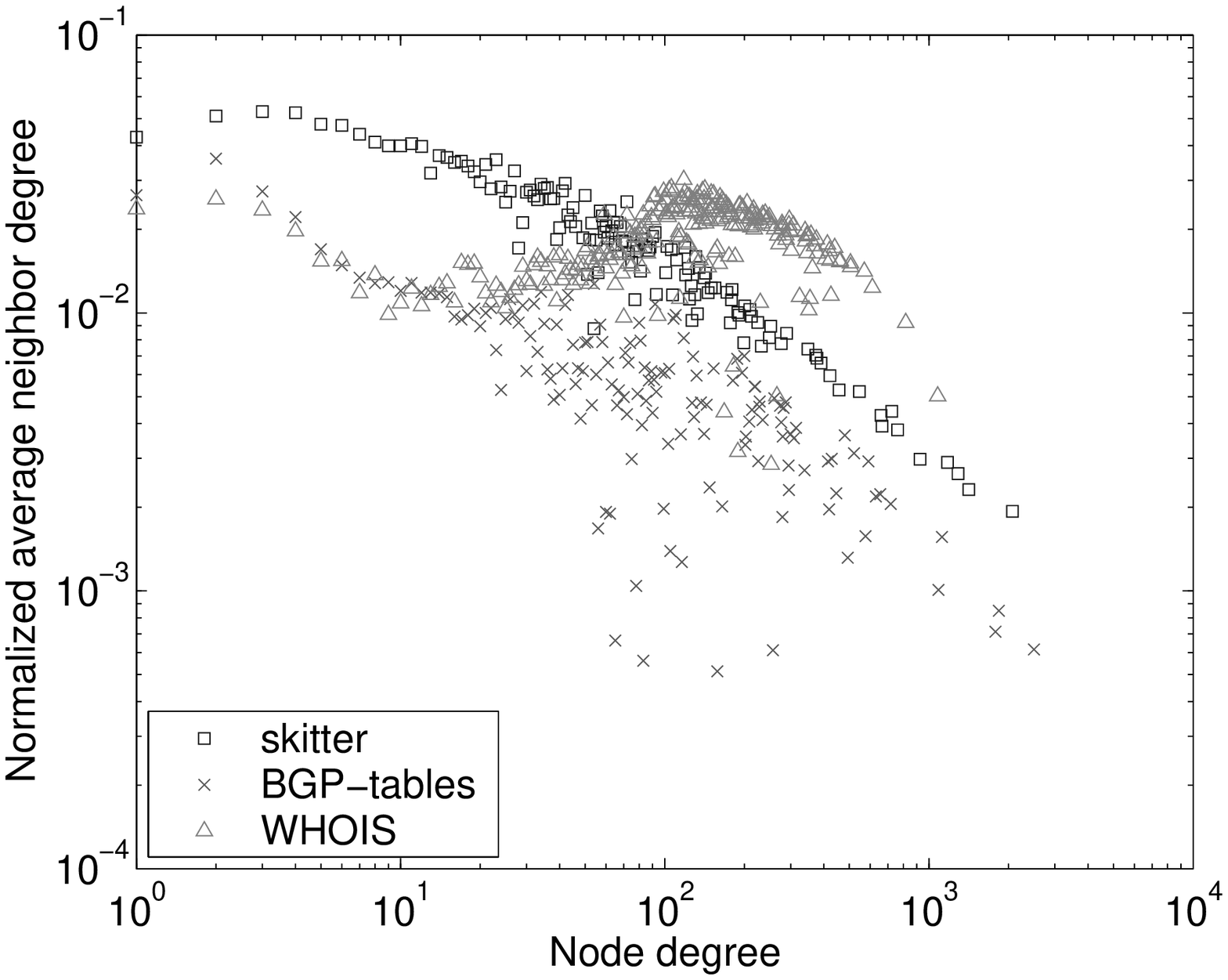}
      }
      \caption{\footnotesize \bf Normalized average neighbor
        connectivity~$\mathbf{k_{nn}(k)/(n-1)}$.
      }
      \label{fig:knnk}
  \end{minipage}
  \hfill
  \begin{minipage}[t]{2.2in}
      \centerline{
          \includegraphics[width=2.2in]{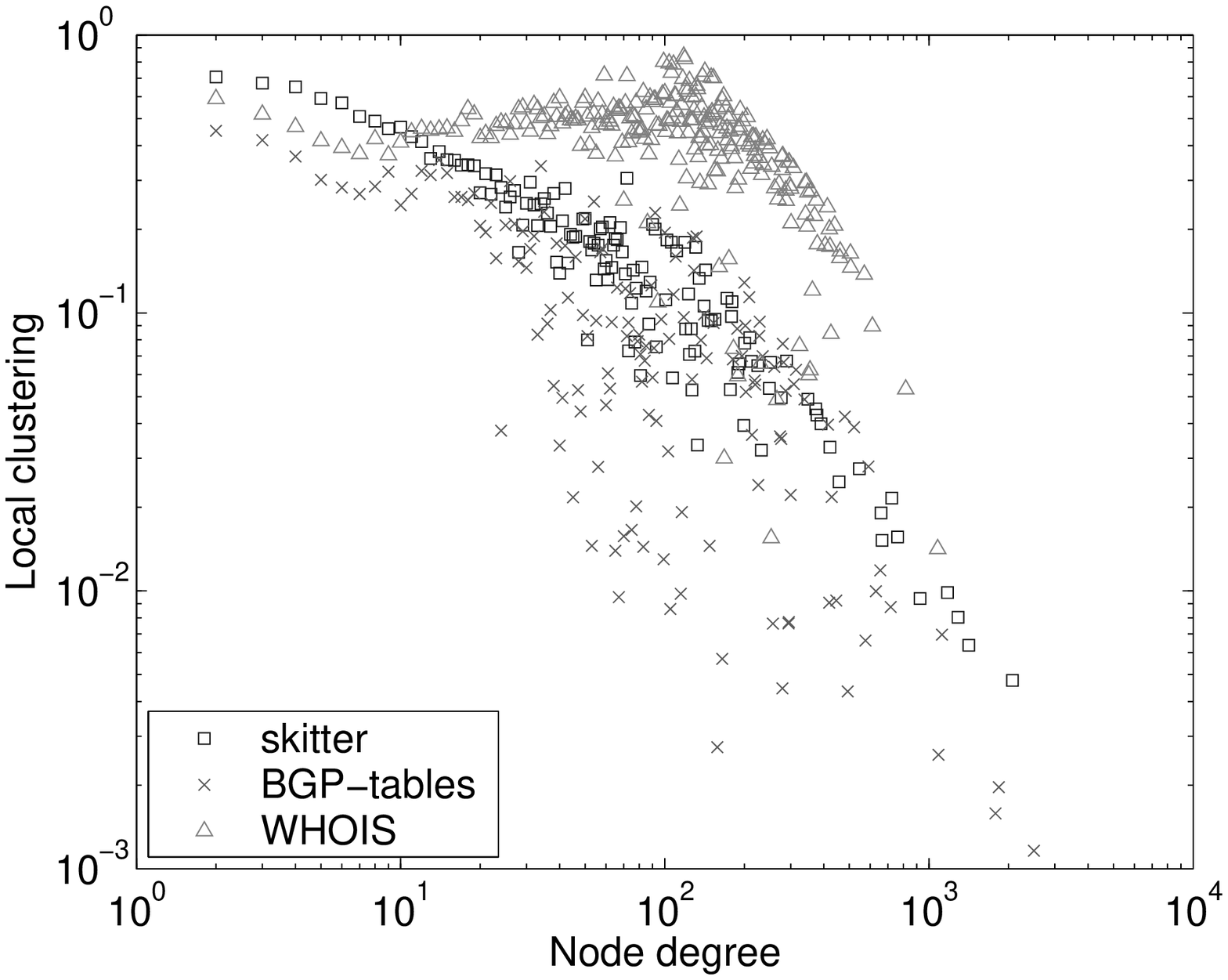}
      }
      \caption{\footnotesize \bf Local clustering~$\mathbf{C(k)}$.
      }
      \label{fig:ck}
  \end{minipage}
\hfill
  \begin{minipage}[t]{2.2in}
      \centerline{
    \includegraphics[width=2.2in]{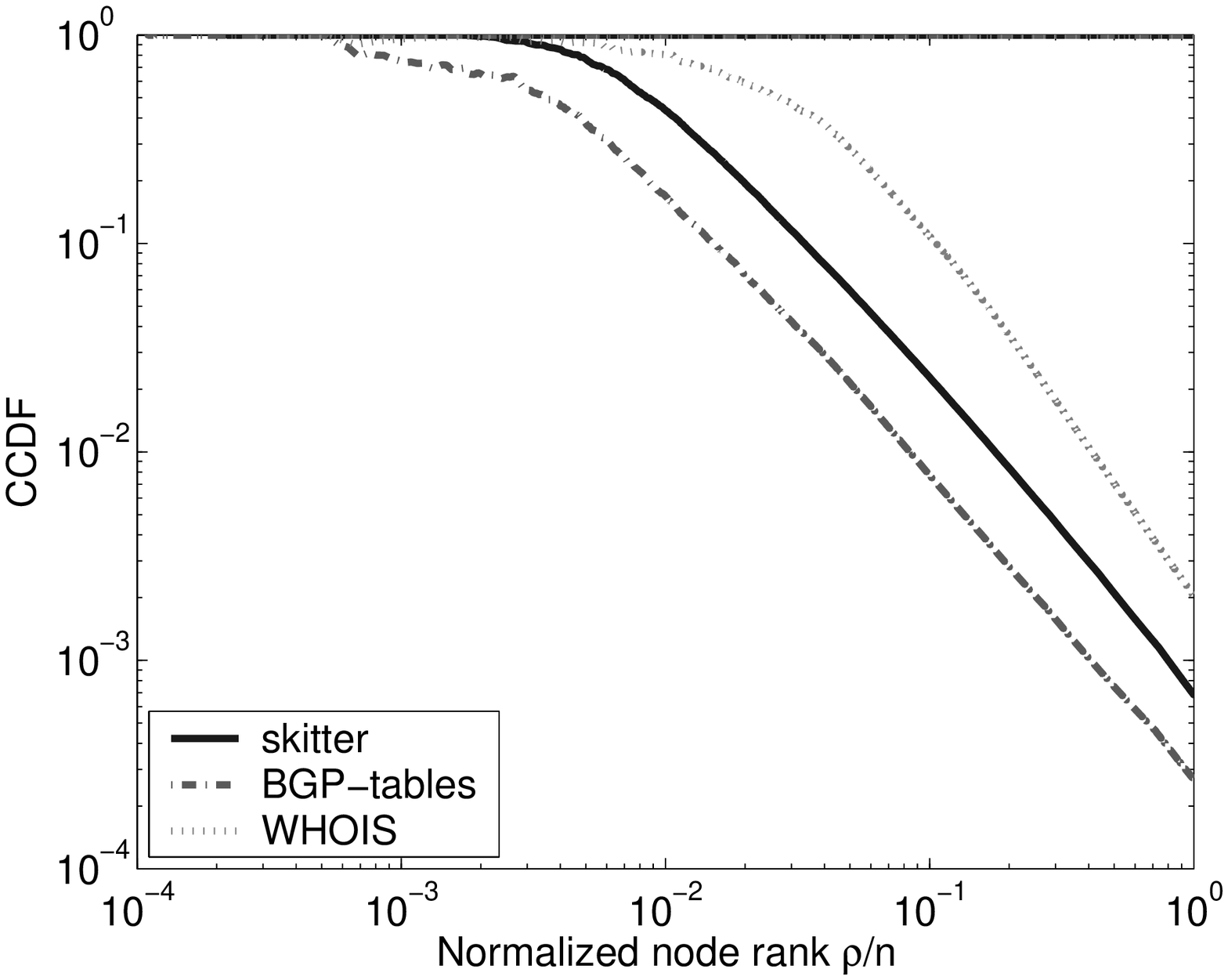}
    }
      \caption{\footnotesize \bf Rich club connectivity~$\mathbf{\phi(\rho/n)}$.}
      \label{fig:rich_club}
 \end{minipage}
\end{figure*}

The interplay between $\bar{k}$- and $r$-orders underlies Figure~\ref{fig:knnk},
where we show the average neighbor connectivity functions for the three graphs.
Skitter has the largest excess of radial links that
connect low-degree nodes (customers ASes) to high-degree nodes (large provider ASes). The high radial links are responsible for skitter's highest average degree for the neighbors of low-degree nodes:
in Figure~\ref{fig:knnk}, skitter is at the top in the area
of low degrees, which follows the $r$-order. On the other hand, the greatest proportion of tangential links
between ASes of similar degrees in WHOIS graph contributes to connectivity of
neighbors of high-degree nodes; therefore the WHOIS graph is at the top for high degree nodes ($\bar{k}$-order).

Note that in the case of skitter and BGP, \mbox{$k_{nn}(k)$}
can be approximated by a power law with the corresponding exponents $\gamma_{nn}$ in Table~\ref{table:summary}.

\subsection{Clustering}\label{sec:clustering}

While the JDD contains information about the degrees of neighbors for
the average $k$-degree node, it does not tell us how
these neighbors interconnect. Clustering satisfies this  need by providing a measure of how close a node's neighbors are to forming a clique.

\textbf{\textit{Definition.}}
Let~\mbox{$\bar{m}_{nn}(k)$} be the average number of
links between the neighbors of $k$-degree nodes.
{\bf Local clustering} is the ratio of this number to
the maximum possible such
links: \mbox{$C(k)=2\bar{m}_{nn}(k)/k/(k-1)$}. If two neighbors of a node are connected, then these three nodes
together form a triangle (3-cycle). Therefore, by definition, local clustering
is the average number of 3-cycles involving $k$-degree nodes. The two summary
statistics associated with local clustering are {\bf mean local clustering}~$\bar{C}=\sum C(k)P(k)$, which is the average value of~$C(k)$, and the {\bf clustering coefficient}~$C$, which is the percentage of 3-cycles among all connected node triplets in the entire graph (for exact definition, see~\cite{BoRi02}).

\textbf{\textit{Importance.}}  Similar to the JDD, one can use
clustering as a litmus test for verifying the accuracy of a topology
model or generator~\cite{BuTo02}. Clustering is a basic connectivity
characteristic. Therefore, if a model reproduces clustering
incorrectly, it is likely to be less accurate for a variety of graph
characteristics.  We use clustering to verify the efficacy of
topology models in Section~\ref{sec:modelcomp}.

Clustering is practical because it expresses local robustness in the graph: the higher the local
clustering of a node, the more interconnected are its neighbors, thus increasing the path diversity locally around the node.
Virus outbreaks spread faster in high-clustered networks, although outbreak sizes are smaller~\cite{newman03b}.
Networks with strong clustering are likely to be
chordal or of low
chordality,\footnote{{\em Chordality\/} of a graph is the length of
the longest cycle without chords. A graph is called {\em chordal\/}
is its chordality is~3.} which makes certain
routing strategies perform better~\cite{fraigniaud05}.

\textbf{\textit{Discussion.}}
We first observe that the clustering average values~$\bar{C}$
in Table~\ref{table:summary} are in the
$\bar{k}$-order, which is expected: more the links, more the clustering. The values of~$\bar{C}$ are almost equal for
skitter and WHOIS, but the clustering coefficient~$C$ is
15 times larger for WHOIS than for skitter. As shown in~\cite{SoVa04},
orders of magnitude difference between~$\bar{C}$ and~$C$ is intrinsic
to highly disassortative networks and is a consequence of
degree correlations~(JDD).

Similarly to~$k_{nn}(k)$,
the interplay between $\bar{k}$- and $r$-orders explains Figure~\ref{fig:ck},
where we plot
local clustering as a function of node degree~$C(k)$. For low degree nodes, skitter's clustering is the highest amongst the three graphs because skitter graph is most disassortative. The links adjacent to low-degree nodes are most likely to lead to high-degree nodes, the latter being interconnected with a high probability. For high degree nodes, the WHOIS graph exhibits highest values for clustering since this graph has the highest average connectivity
(largest~$\bar{k}$). The neighbors of high-degree nodes are interconnected to a greater extent resulting in higher clustering for such nodes.

Similar to~$k_{nn}(k)$, $C(k)$ also can be
approximated by a power law for skitter and BGP graphs (exponents~$\gamma_C$
in Table~\ref{table:summary}).

JDDs with strong correlations play a major part for the presence of non-trivial clustering observed in
many networks~\cite{SoVa04}. This interplay explains
overall similarity between degree correlations and clustering, in general,
and similarity between~$k_{nn}(k)$ and~$C(k)$, in particular.

\subsection{Rich club connectivity}

\textbf{\textit{Definition.}}
Let~\mbox{$\rho = 1 \ldots n$} be the first~$\rho$ nodes ordered by
their non-increasing degrees in a graph of size~$n$. {\bf Rich club
connectivity}~(RCC)~$\phi(\rho/n)$ is the ratio of the number of links in the subgraph
induced by the~$\rho$ largest-degree nodes to the maximum possible links~\mbox{$\rho(\rho-1)/2$}. In other words, the RCC is a measure of how
close $\rho$-induced subgraphs are to cliques.

\begin{figure}[tbh]
      \centerline{
          \includegraphics[width=2.75in]{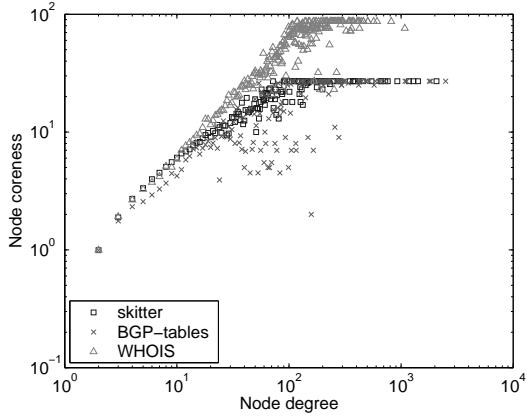}
      }
      \caption{\footnotesize \bf Average coreness of $k$-degree nodes~$\mathbf{\kappa(k)}$.}
      \label{fig:kappak}
\end{figure}

\textbf{\textit{Importance.}}
As of this writing, one of the more successful Internet AS-level topology model is the Positive Feedback Preference~(PFP) model by Zhou and Mondragon~\cite{ZhoMo04}.
It accurately reproduces a wide spectrum of metrics
of the measured AS-level topology by trying to explicitly capture only
the following three characteristics: (i)~the exact form of the node degree distribution; (ii)~the maximum node degree; and
(iii)~RCC. The success of the PFP model in approximating the real topology
is yet to be fully explained. One can show that networks with the same
{\em JDDs\/} have the same RCC. The converse is not true, but one can fully describe
all the JDDs having a given form of RCC.

\textbf{\textit{Discussion.}}
As expected, the values of~\mbox{$\phi(\rho/n)$} in Figure~\ref{fig:rich_club}
are in the $\bar{k}$-order with WHOIS at the top: more links result in denser cliques. RCC exhibits clean power laws for all three graphs in the area of medium and large~\mbox{$\rho/n$}. The values of the power-law exponents~$\gamma_{rc}$ in Table~\ref{table:summary} result from fitting~$\phi(\rho/n)$ with power laws
for~90\% of the nodes, \mbox{$0.1 \leq \rho/n \leq 1$}.

\subsection{Coreness}\label{sec:coreness}

\textbf{\textit{Definition.}} There are two definitions of
coreness. In graph-theoretic literature~\cite{bollobas85}, the
$k$-core of a graph is the subgraph obtained from the original
graph after removal of all nodes of degree less than or equal
to~$k$. A more informative definition of 
$k$-core~\cite{GaPa04} is the subgraph obtained from the
original graph by the {\em iterative\/} removal of all nodes of degree less
than or equal to~$k$.\footnote{Remove all nodes of degree~$\leq
k$, then do it again in the remaining graph, proceed until all
remaining nodes are of degrees~$>k$.} We use the latter
definition. The {\bf node coreness}~$\kappa$ of a given node is
then the maximum~$k$ such that this node is still present in the
$k$-core but removed in the \mbox{$(k+1)$}-core. The minimum node
coreness in a given graph is~\mbox{$\kappa_{min}=k_{min}-1$},
where $k_{min}$ is the lowest node degree present. All 1-degree
nodes have $\kappa=0$. The maximum node coreness~$\kappa_{max}$ in
a graph, or the {\bf graph coreness}, is such that the
$\kappa_{max}$-core is not empty, but $(\kappa_{max}+1)$-core is.
For example, coreness of a tree is~$0$ and coreness of a
$k$-regular graph~\cite{harary94} is equal to coreness of all of
its nodes (all having degree~$k$), which is~\mbox{$k-1$}. We
further define the {\bf graph core} as its $\kappa_{max}$-core,
and the {\bf graph fringe} as the set of nodes with minimum
coreness~$\kappa_{min}$. Note that because the process of building
core is iterative, nodes with degree $k>\kappa_{min}$ can be in
the fringe.

\textbf{\textit{Importance.}} The node coreness tells us how
``deep in the core'' the node is. It is a much more sophisticated
measure of node connectivity than node degree. Indeed, the node
degree can be high, but if its coreness is small, then the node is
not well connected and one can easily disconnect it by removing
its poorly connected neighbors. For example, a high-degree hub
of a star has coreness of~$0$. At the same time, node coreness is
not a measure of centrality of the node. For example, a low-degree
node interconnecting a few high-degree hubs has a low value of
coreness, but intuitively it is in the ``center of the graph.'' At
the same time, coreness is important for topology visualization
capable of revealing network architectural
fingerprints~\cite{AlDaBaVe04} and signatures of topology dynamics
under different types of anomalies (worm and DoS attacks, outages,
misconfigurations, etc.)~\cite{GaPa04}.

\textbf{\textit{Discussion.}} The average node coreness in
Table~\ref{table:summary} is in the $\bar{k}$-order, which is
expected. The graph coreness of WHOIS is more that three times
larger than of skitter and BGP. WHOIS has particularly large core
size and graph coreness because the $r$-order amplifies the
$\bar{k}$-order in this case: WHOIS has highest link density
(largest~$\bar{k}$) and highest concentration of them in the core
(largest~$r$). WHOIS graph has the largest relative core size and
smallest relative fringe size (cf.~Table~\ref{table:summary}). The
BGP graph is the sparsest, having the smallest relative core size
and the largest relative fringe size. Interestingly, in the BGP
graph, nodes with degree as low as~34 are in the core, and nodes
with degree as high as~7 are in the fringe. For all three graphs,
the average node coreness as a function of node
degree~\mbox{$\kappa(k)$} roughly follows power laws
for~\mbox{$k\lesssim100$} (Figure~\ref{fig:kappak}). The
corresponding exponents and mean coreness are given in
Table~\ref{table:summary}. For nodes with
degrees~\mbox{$k\gtrsim100$} the coreness reaches saturation:
increasing node degree above~$100$ does not increase coreness.

\subsection{Distance}\label{sec:characteristics:distance}

\textbf{\textit{Definition.}}
The shortest path length distribution or
simply the {\bf distance
distribution}~\mbox{$d(x)$} is the probability for a random pair of nodes to
be at a distance~$x$ hops from each other. Two basic summary statistics
associated with the distance distribution of a graph are {\bf average
distance}~$\bar{d}$ and the {\bf standard deviation}~$\sigma$. We call the
latter the {\em distance distribution width} since distance distribution in
 Internet graphs (and in many other networks) has a characteristic
Gaussian-like shape.

Eccentricity is an extreme form of distance: if~$d_{ij}$ is
distance between nodes~$i$ and~$j$, then
{\bf eccentricity}~$\varepsilon_i$ of node~$i$
is the maximum distance from~$i$~\cite{harary94}:
\mbox{$\varepsilon_i = \max_j d_{ij}$}.
The maximum eccentricity in a graph is also the maximum distance and is called
the graph {\bf diameter}~$D=\varepsilon_{max}$, and the minimum
eccentricity~$R=\varepsilon_{min}$ is called the graph {\bf radius}. The set
of nodes with maximum eccentricity forms {\bf graph periphery}, while nodes
with minimum eccentricity belong to {\bf graph center}~\cite{harary94}.

\begin{figure*}[tbh]
  \begin{minipage}[t]{2.2in}
      \centerline{
          \includegraphics[width=2.25in]{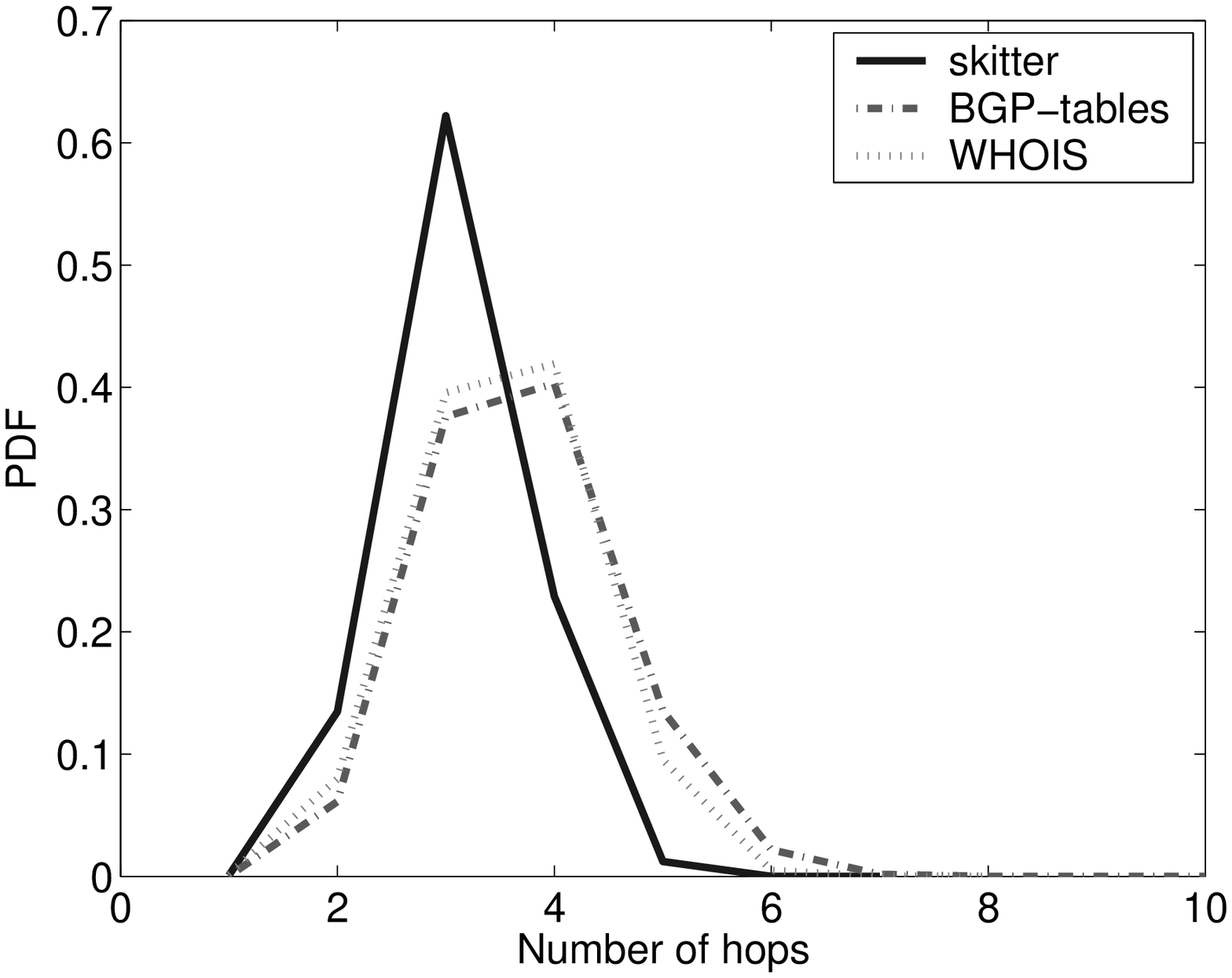}
      }
      \caption{\footnotesize \bf Distance~$\mathbf{d(x)}$ distribution.
    }
      \label{fig:distance-pdf}
  \end{minipage}
  \hfill
  \begin{minipage}[t]{2.2in}
      \centerline{
          \includegraphics[width=2.25in]{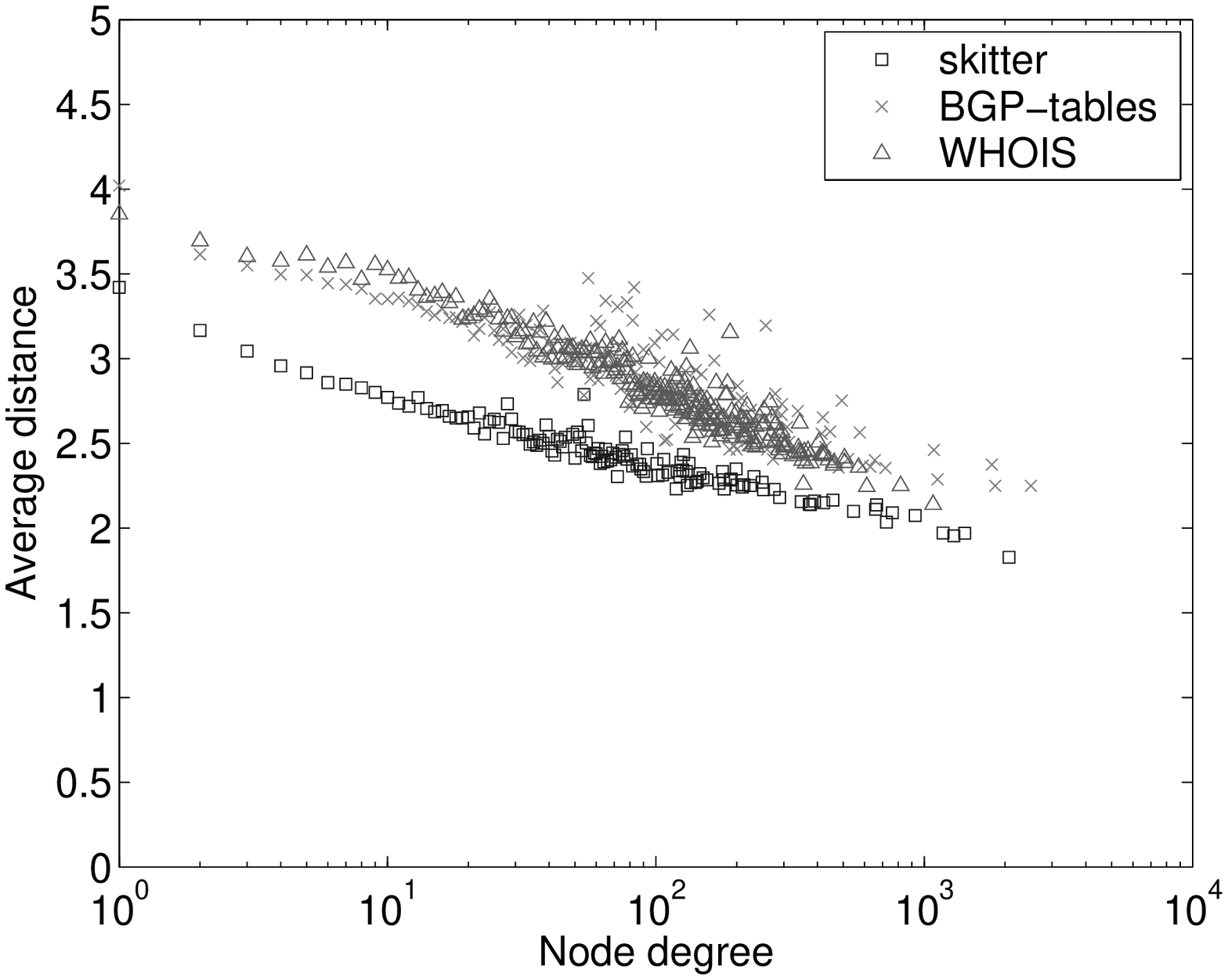}
      }
      \caption{\footnotesize \bf Average distance from $k$-degree nodes~$\mathbf{d(k)}$.}
      \label{fig:distance-k}
  \end{minipage}
  \hfill
 \begin{minipage}[t]{2.2in}
      \centerline{
          \includegraphics[width=2.25in]{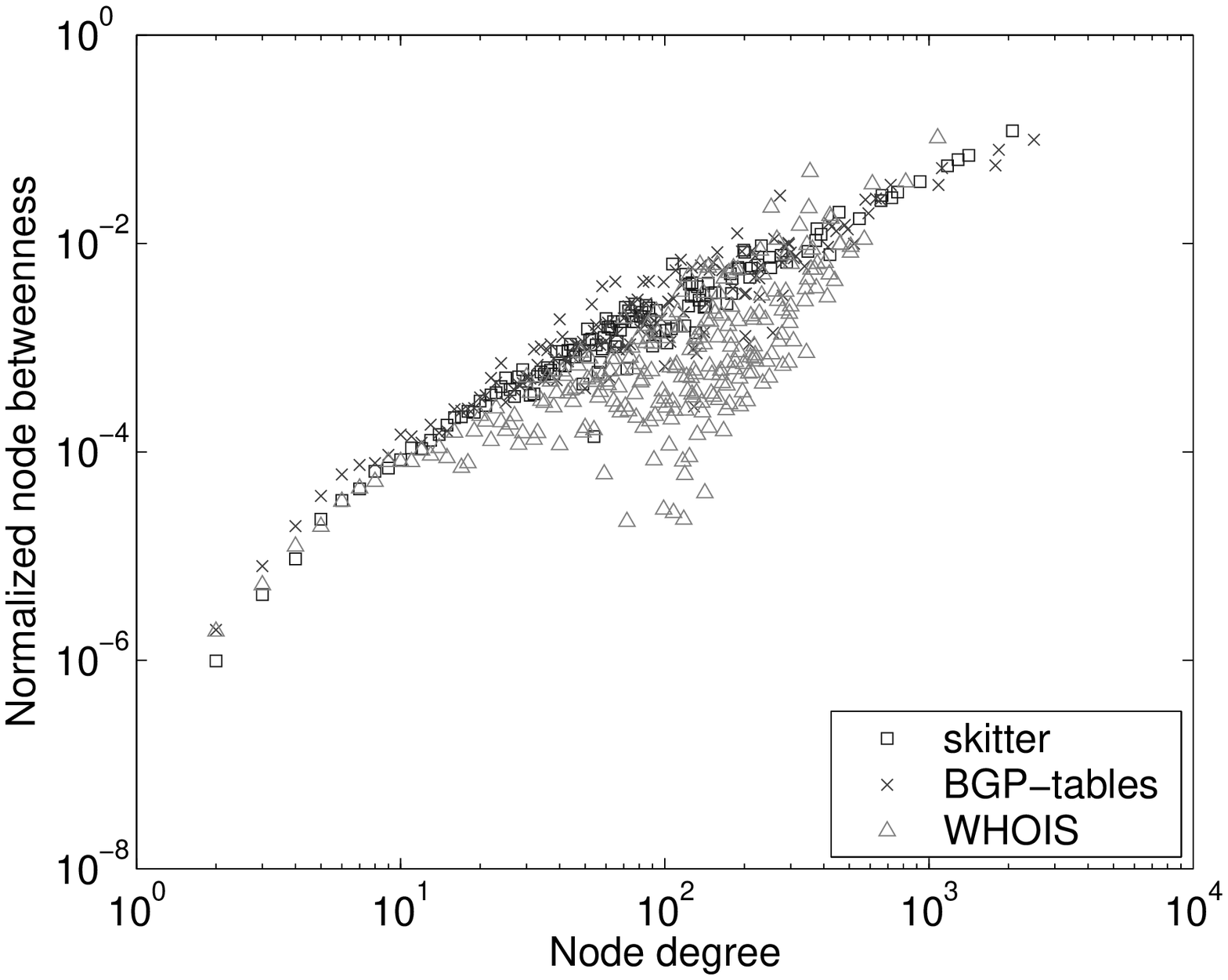}
      }
      \caption{\footnotesize \bf Normalized node betweenness~$\mathbf{B(k)/n/(n-1)}$.
}
      \label{fig:betweenness}
  \end{minipage}
  \hfill
\end{figure*}

\begin{figure}[tbh]
	\centerline{
          \includegraphics[width=2.75in]{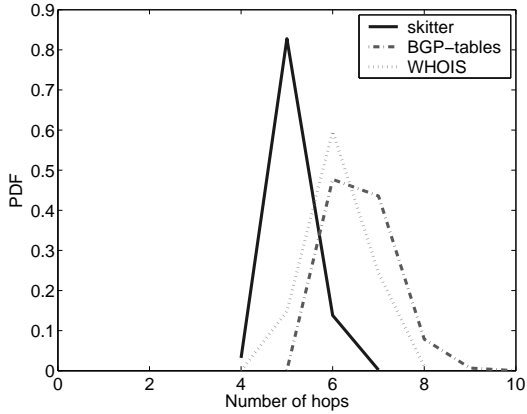}
	}
      \caption{\footnotesize \bf Eccentricity $\mathbf{\varepsilon(k)}$ distribution.}
      \label{fig:ecc-pdf}
\end{figure}

\begin{figure}[tbh]
	\centerline{
          \includegraphics[width=2.75in]{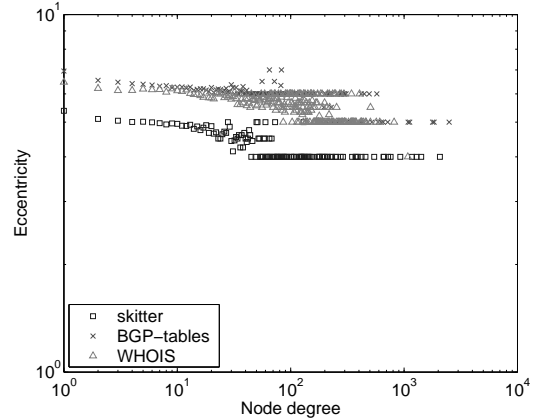}
	}
      \caption{\footnotesize \bf Average eccentricity from $k$-degree nodes~$\mathbf{ \varepsilon(k)}$.}
      \label{fig:ecc-k}
\end{figure}

\textbf{\textit{Importance.}}
Distance distribution is critically important for many applications,
the most prominent
being routing. Distance-based locality-sensitive approach~\cite{peleg01}
is the root of most modern routing algorithms. As shown
in~\cite{KrFaYa04}, performance parameters of these algorithms depend
strongly on the distance distribution in a network. In particular,
short average
distance and narrow distance distribution width break the efficiency of
traditional hierarchical routing. They are among the root causes
of interdomain routing scalability issues in the Internet today.

Distance distribution also plays a vital role in robustness of the network to viruses. Viruses
can quickly contaminate larger portions of a network that has
small distances between nodes. Topology models that accurately reproduce observed distance distribution will benefit researchers, who are developing techniques to quarantine the network from viruses.
Finally, {\em expansion\/}
from the seminal paper~\cite{TaGoJaShWi02}, identified as a critical metric
for topology comparison analysis, is a renormalized version
of distance distribution.

\textbf{\textit{Discussion.}}
Interestingly enough, although the distance distribution is a
``global'' topology characteristic, we can explain Figure~\ref{fig:distance-pdf} by the interplay between our local connectivity characteristics: the
$\bar{k}$- and $r$-orders. First, we note
that the skitter graph stands out in Figure~\ref{fig:distance-pdf} as it has the
smallest average distance and the smallest distribution width
(cf.~Table~\ref{table:summary}). This result appears unexpected at first since the
skitter graph has more nodes than the WHOIS graph and only about half the number of
links. One would expect a denser graph (WHOIS) to have a lower average distance
since adding links to a graph can only {\em decrease\/} the
average distance in it. Surprisingly, the average distance of the most richly
connected (highest~$\bar{k}$) WHOIS graph is not the lowest. This result can be explained using
the $r$-order.
Indeed, a more disassortative graph has a greater proportion of radial links,
shortening the distance from the fringe to the core.\footnote{Henceforth, we
use terms {\em fringe} and {\em core} to mean zones in the graph with low-
and high-degree nodes respectively.}
The skitter graph has the right balance between the
relative number of links~$\bar{k}$ and their radiality~$r$,
that minimizes the average distance. Compared to skitter, the BGP graph has larger
distance because it is sparser
(lower~$\bar{k}$), and the WHOIS graph has larger distance because it is more
assortative (higher~$r$).

The fact that~62\% of AS paths in the skitter graph are 3-hop paths suggests
the most frequent path pattern reflecting the customer-provider AS hierarchy:
source's AS in the fringe~$\rightarrow$ source's
provider AS in the core~$\rightarrow$ destination's provider AS in the
core~$\rightarrow$ destination's AS in the fringe.

Another important observation is that for all three graphs, including WHOIS,
the average distance as a function of node degree exhibits relatively stable
power laws in the full range of node degrees (Figure~\ref{fig:distance-k}),
with exponents given in Table~\ref{table:summary}.

Both the eccentricity distribution~\mbox{$\varepsilon(x)$}
(Figure~\ref{fig:ecc-pdf}) and average eccentricity from
$k$-degree nodes~\mbox{$\varepsilon(k)$} (Figure~\ref{fig:ecc-k}) are similar
to their averaged distance counterparts.
Table~\ref{table:summary} also shows diameter, radius and average eccentricity
for our graphs, as well as the relative size of graph center and
periphery. In the WHOIS graph, the center consists of only one AS, AS702
(UUNET), uniquely positioned to have the minimum eccentricity of~$4$. If we
add the nodes having eccentricity of~$5$, the center would consist of
1109~ASs, the center size ratio~$n_R/n=0.15$ would become the largest
among all three graphs, and it would be in the expected $\bar{k}$-order.

\subsection{Betweenness}\label{sec:characteristics:betweenness}

Average distance is a  good node centrality measure: intuitively, nodes with smaller
average distances are closer to the graph ``center.'' However, the most
commonly used measure of centrality is betweenness. It is applicable not
only to nodes, but also to links.

\textbf{\textit{Definition.}}
Let~$\sigma_{ij}$ be the number
of shortest paths between nodes~$i$ and~$j$ and~$l$ be either a node
{\em or\/} link. Let~$\sigma_{ij}(l)$ be the number of shortest
paths between~$i$ and~$j$ going through node (or link)~$l$. Its
{\bf betweenness} is~\mbox{$B_l = \sum_{ij}\sigma_{ij}(l)/\sigma_{ij}$}.
The maximum possible value for node and link betweenness is $n(n-1)$~\cite{DaAlHaBaVaVe05},
therefore in order to compare betweenness in graphs of different sizes, we
normalize it by $n(n-1)$.

\begin{figure*}[tbh]
    \centerline{
        \subfigure[skitter]
        {\includegraphics[width=2in]{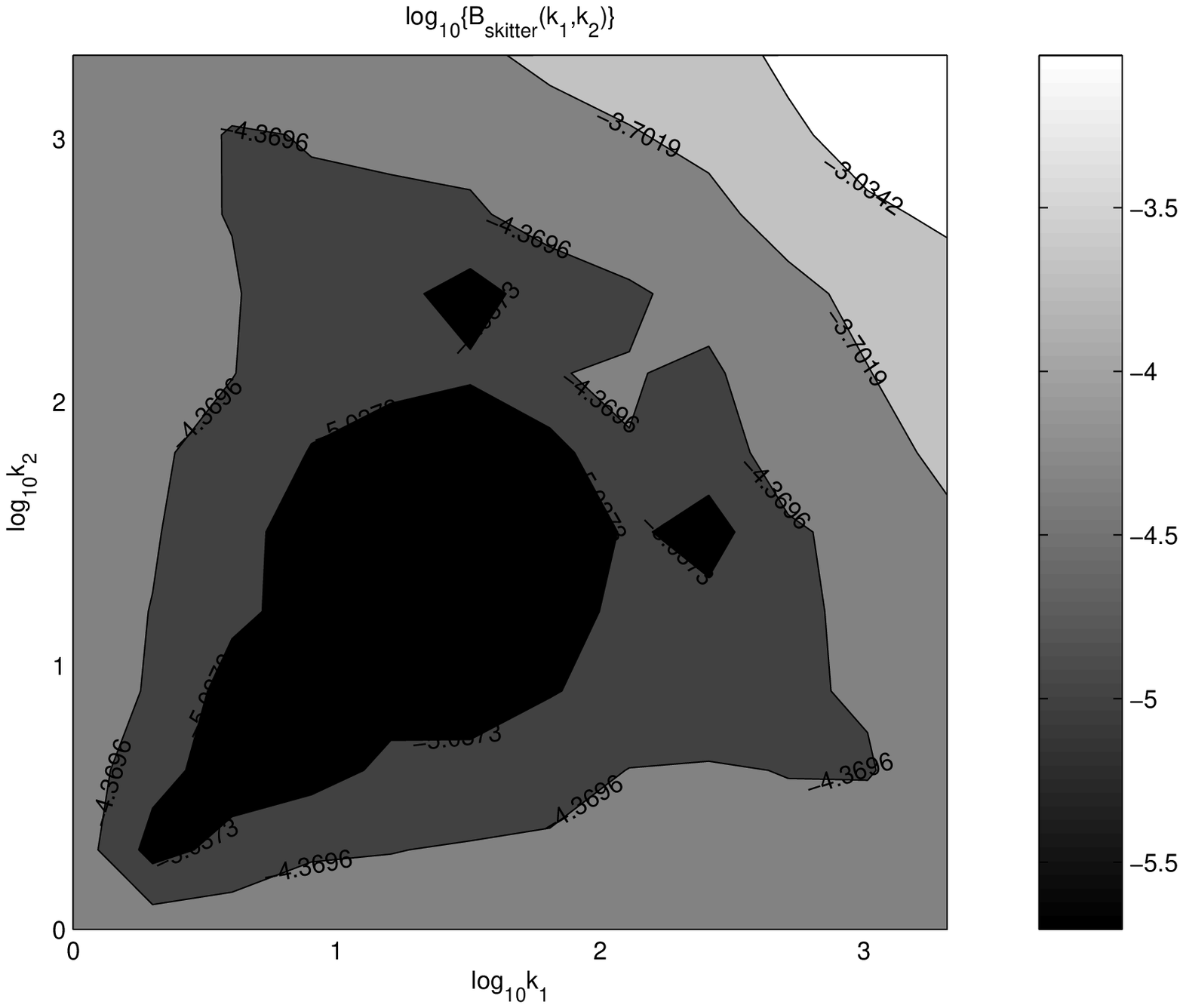}
        \label{fig:bkk-skitter}}
        \hfill
        \subfigure[BGP tables]
        {\includegraphics[width=2in]{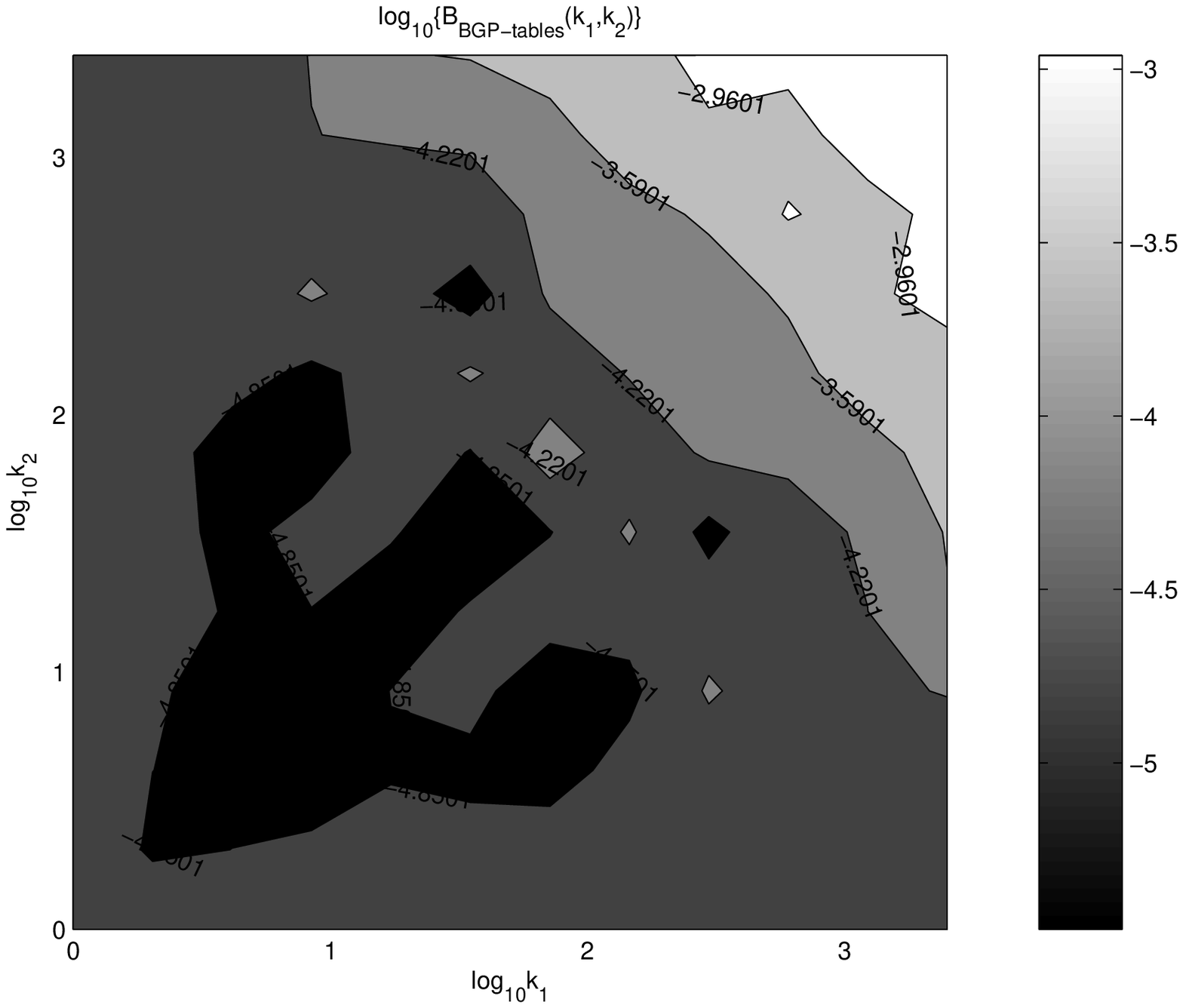}
        \label{fig:bkk-bgp}}
        \hfill
        \subfigure[WHOIS]
        {\includegraphics[width=2in]{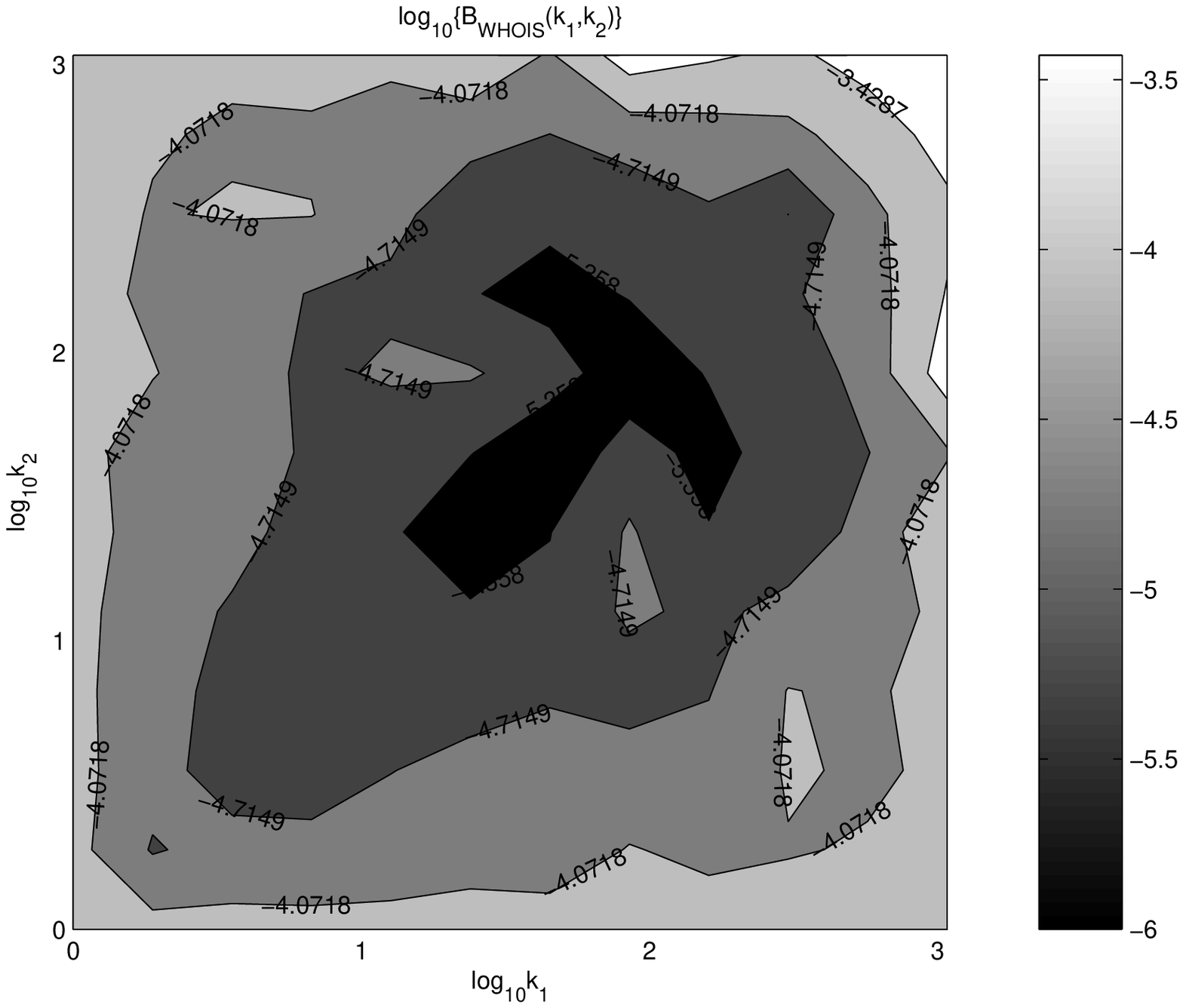}
        \label{fig:bkk-whois}}
    }
    \caption{\footnotesize {\bf Logarithm of normalized link
    betweenness~$\mathbf{B(k_1,k_2)/n/(n-1)}$ on a log-log scale.}}
    \label{fig:bkk}
\end{figure*}

\textbf{\textit{Importance.}}
Betweenness measures the number of shortest paths
passing through a node or link and, thus, estimates the potential traffic load
on this node/link assuming uniformly distributed traffic following shortest
paths.\footnote{In fact, some variants of betweenness are just called
{\em load\/}~\cite{brandes01}.}
Betweenness is important for traffic engineering applications that
try to estimate potential traffic load on nodes/links and potential congestion
points in a given topology. Betweenness is also critical for evaluating the
accuracy of topology sampling by traceroute-like probes (e.g.~skitter and BGP).
As shown in~\cite{DaAlHaBaVaVe05}, the
broader the betweenness distribution, the higher the statistical accuracy of the
sampled graph. The exploration process statistically focuses on
nodes/links with high betweenness thus providing an accurate sampling of the distribution tail and
capturing relevant statistical information. Finally we note that
{\em link value}, used in~\cite{TaGoJaShWi02} to analyze the topology hierarchy,
and {\em router utilization}, used~\cite{LiAlWiDo04} to measure network
performance, are both directly related to betweenness.

\textbf{\textit{Discussion.}}
The simplest approach to calculating node betweenness results in long running
times, but we used an efficient algorithm from~\cite{brandes01}. We also modified it
to also compute link betweenness. For skitter and BGP graphs,
node betweenness is a growing power-law function of node degree
(Figure~\ref{fig:betweenness}) with exponents given in Table~\ref{table:summary}.
The WHOIS graph has an excess of medium degree nodes (cf.~Figure~\ref{fig:pk})
leading to greater path diversity and, hence, to lower betweenness values for
these nodes.
We also calculate average link betweenness as a function of degrees
of nodes adjacent to a link~\mbox{$B(k_1,k_2)$} (Figure~\ref{fig:bkk}).
Contrary to popular belief, the contour plots show that link betweenness does not measure link centrality.
First, betweenness of links adjacent to low-degree nodes
(the left and bottom sides of the plots) is not the minimum. In fact,
non-normalized betweenness of links adjacent to 1-degree nodes is constant and equal
to~\mbox{$n-1$} (the number of destinations in the rest of the network).
Similar values of betweenness characterize links elsewhere
in the graph, including radial links between high and low-to-medium degree
nodes and tangential links in the zone of medium-to-high degrees (diagonal zone
from bottom-right to upper-left). Second, while the maximum-betweenness links
are between high-degree nodes as expected (the upper right corner of the plots),
the minimum-betweenness links are tangential in the medium-to-low degree zone
(diagonal areas of low values from bottom-left to upper-right). We can explain
the latter observation by the following argument. Let~$i$ and~$j$ be two nodes
connected by a minimum-betweenness link~$l$. The only shortest paths going
through~$l$ are those between nodes that are {\em below}~$i$ and~$j$, where
``below'' means further from the core and closer to the fringe. When the
degrees of both~$i$ and~$j$ are small, the numbers of nodes below them (with
lower degree) are small, too. Consequently, the number of shortest paths,
proportional to the product of the number of nodes below~$i$ and~$j$, attains
its minimum at~$l$. We conclude that link betweenness is not a measure of
centrality but a measure of some combination of link centrality and radiality.

\subsection{Spectrum}\label{sec:characteristics:spectrum}

\textbf{\textit{Definition.}}
Let~$\hat{a}$ be the adjacency matrix of a graph. This~\mbox{$n \times n$}
matrix is constructed by setting the value of its
element~\mbox{$a_{ij}=a_{ji}=1$} if there is a link between nodes~$i$ and~$j$.
All other elements have value ~$0$. Scalar~$\lambda$ and vector~$v$ are
the eigenvalue and eigenvector respectively of~$\hat{a}$ if~\mbox{$\hat{a} v = \lambda v$}.
The {\bf spectrum} of a graph is the set of eigenvalues of its
adjacency matrix.

\begin{figure}[tbh]
      \centerline{
          \includegraphics[width=2.75in]{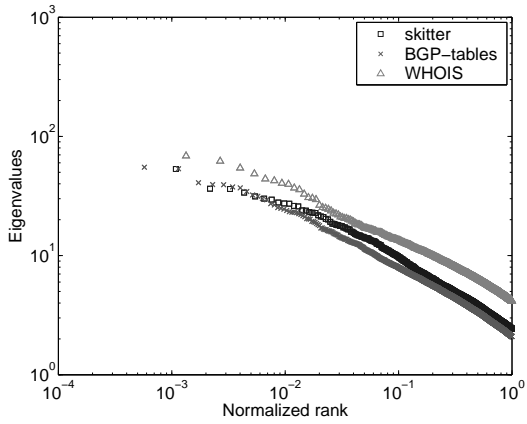}
      }
      \caption{\footnotesize {\bf Spectrum.} Absolute values of top 10\% of
      eigenvalues ordered by their normalized rank: the absolute value divided
      by the total number of eigenvalues calculated for a given graph.}
      \label{fig:spectrum}
\end{figure}
\textbf{\textit{Importance.}}
We stress that spectrum is one of the most important
{\em global\/} characteristics of the topology.
Spectrum yields tight bounds for a {\em wide range\/} of critical graph
characteristics~\cite{chung97}, such as distance-related parameters, expansion
properties, and values related to separator problems estimating graph resilience
under node/link removal. The largest eigenvalues are particularly important.
Most networks with high largest eigenvalues have small
diameter, expand faster, and are more robust. To further emphasize the importance of
spectrum, we consider the following two specific examples of spectrum-related
metrics that played a central role in two significant contributions to networking
topology research.

First, Tangmunarunkit {\it et al.}~\cite{TaGoJaShWi02} defined network {\em resilience},
one of the three metrics critical for their topology comparison analysis, as a measure
of network robustness under link removal, which equals the minimum balanced cut size of
a graph. By this definition, resilience is related to spectrum since the graph's largest
eigenvalues provide bounds on network robustness with
respect to both link {\em and\/} node removals~\cite{chung97}.

Second, Li {\it et al.}~\cite{LiAlWiDo04} define network {\em performance}, one of the two
metrics critical for their HOT argument, as the maximum traffic throughput of the network. By this definition, performance is related to spectrum since it is essentially the
network conductance~\cite{GkaMiSa03}. It can be tightly estimated by the gap between the first and second
largest eigenvalues~\cite{chung97}.

Beyond its significance for network robustness and performance,
the graph's largest eigenvalues are important
for traffic engineering purposes since graphs with larger eigenvalues
have, in general, more node- and link-disjoint paths to choose from.
The spectral analysis of graphs is also a powerful tool for detailed investigation
of network structure ~\cite{VuHuEr01,GkaMiZe03}
, such as discovering clusters
of highly interconnected nodes, and can reveal the hierarchy of ASes in the
Internet~\cite{GkaMiZe03}.

\textbf{\textit{Discussion.}}
Our $\bar{k}$-order (BGP, skitter, WHOIS)
plays a key role once again: the densest graph, WHOIS is on the top in
Figure~\ref{fig:spectrum} and its first eigenvalue is largest in Table~\ref{table:summary}.
The eigenvalue distributions of all the three graphs follow power laws.

Other important metrics such as coreness and eccentricity are explained in detail in the Supplement~\cite{comp-anal}. As with other metrics, the resulting metric values and differences in the three data sources can be explained using $\bar{k}$-order and $r$-order.

\section{Observed topologies vs.~random graph models}
\label{sec:modelcomp}

So far we have looked at metrics that provide important details about
the Internet AS-graph. These metrics directly impact network
applications and protocols, and can also be used to distinguish
between different topologies. Using JDD, which determines both
$\bar{k}$-order and $r$-order, we have been able to account for the
differences and peculiarities in our target data sets. We next
consider models that aim to reproduce observed topologies. In this
section, we consider different classes of random graphs and discuss
the relationship between these theoretical models and the Internet
graphs we constructed from measurements. This analysis will help
determine how close random graph models come to capturing measured
Internet topologies.

\subsection{Random graph models}\label{sec:deftopo}

Topology generators and models have been evolving steadily in the past few years. The simplest model mimicked the average degree observed in the topology. Given the number of nodes~$n$ and edges~$m$ (e.g.~in the original graph), and,
consequently, the average degree~\mbox{$\bar{k}=2m/n$}, one can construct the class of
maximally random graphs having the same average degree~$\bar{k}$ by connecting
every pair of nodes with probability~\mbox{$p=\bar{k}/n$}. These graphs belong to the class of classical (Erd\H{o}s-R\'{e}nyi) random graphs~$G_{n,p}$~\cite{ErRe59}.
In this paper we call such graphs {\bf 0K-random} conceptualizing them as a zero-order approximation to the connectivity in the original graph.
(We explain the exact semantics behind this terminology at the end of this section.)
In general, 0K-random graphs fail to approximate real Internet
topologies. In particular, the node degree
distribution in 0K-random graphs is binomial, which is closely approximated by Poisson
distribution
\mbox{$P_{0K}(k) = e^{-\bar{k}}\bar{k}^k/k!$}~\cite{DorMen-book03}. It is different from power-law degree distributions observed in the Internet.

The next model remedied this deficiency by capturing the degree distribution of the nodes. Given a specific form of the degree distribution~$P(k)$ (e.g.~extracted from
the original graph), one can construct the class of maximally random graphs
having the same degree distribution following, for example, a recipe introduced
in~\cite{MolRee95,MolRee98} and further formalized in~\cite{AiChLu00}. We call
such graphs {\bf 1K-random}, and we can think of them as providing the
first-order approximation to the connectivity of the original graph.
Of particular interest for Internet modeling is the case when~$P(k)$ is a power-law
function~\cite{FaFaFa99}. The resulting sub-class of 1K-random graphs is called
{\bf power-law random graphs} (PLRG). Note that the 1K-random graphs
have a specific form of the JDD~$P(k_1,k_2)$~\cite{DorMen-book03}.
If we denote by~\mbox{$\tilde{P}(k)$} the probability that one of the two nodes
adjacent to a randomly selected edge is
of degree~$k$, \mbox{$\tilde{P}(k)=(k/\bar{k})P(k)$}, then the JDD in 1K-random
graphs is~\mbox{$P_{1K}(k_1,k_2)=\tilde{P}(k_1)\tilde{P}(k_2)$}, meaning that
there is no correlation between degrees of adjacent nodes. This is why
1K-random graphs are also called {\em uncorrelated} graphs. By construction~\cite{ErRe59}, 0K-random graphs are also uncorrelated,
with their JDD ~$P_{0K}(k_1,k_2)$ given by the same expression as above, where~$P(k)$ is
the Poisson distribution~$P_{0K}(k)$.

We now define a model that provides the next level of approximation: {\bf 2K-random} graphs, which are maximally random graphs reproducing the given JDD~$P(k_1,k_2)$. These graphs have the exact JDD as the original topology, but are random in all other respects.
The semantics behind the ``dK-random'' notation becomes clear now: $d$ in
``dK-random'' is the number of arguments in the degree distribution
function~$P(k_1,k_2,\ldots,k_d)$ that the dK-random graphs reproduce.

\subsection{Comparison with observed topologies}

As demonstrated in~\cite{TaGoJaShWi02}, 1K-random graphs produced by
PLRG-based topology generators produce more accurate
approximations of the Internet topology than outputs of older topology
generators designed to simulate the perceived hierarchical structure
of the Internet. We show that the topology generation strategy based
on modeling only the degree distribution fails to attain the level of
accuracy required in the description of Internet topology. Li {\it et
al.} ~\cite{LiAlWiDo04} have shown that graphs with the same degree
distribution can have different structures. In section~\ref{sec:jddmodel}, we compare the JDD of 1K-random graphs to the JDD
observed in the measured data and show how they are different. As a
next step, we also show how 2K-random graph models better approximate
the real topologies.

\subsubsection{Joint degree distribution}\label{sec:jddmodel}
\begin{figure*}[tbh]
    \centerline{
        \subfigure[1K-random graph with skitter's~$P(k)$]
        {\includegraphics[width=1.7in]{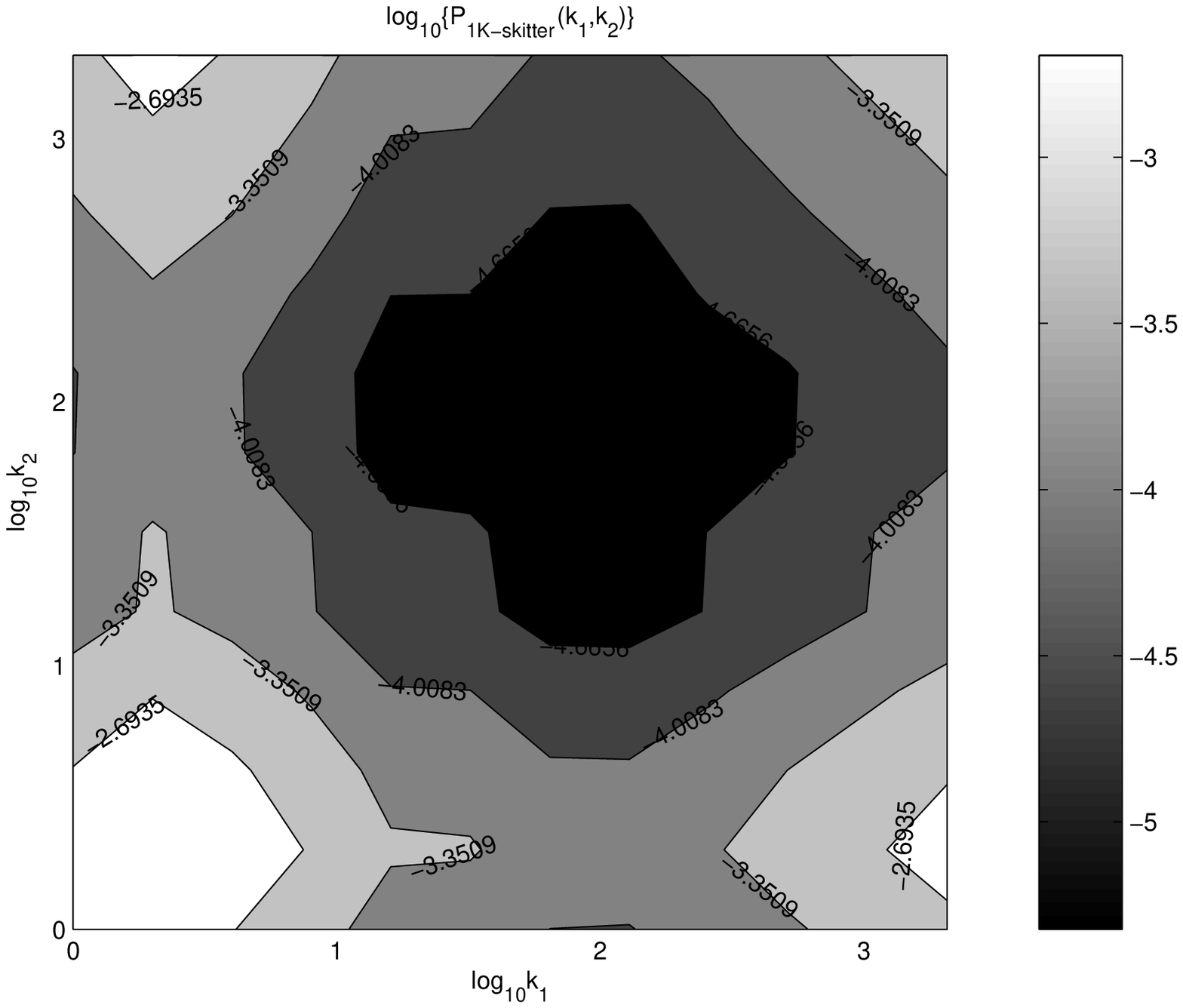}
        \label{fig:ppk-skitter}}
        \hfill
        \subfigure[skitter]
        {\includegraphics[width=1.7in]{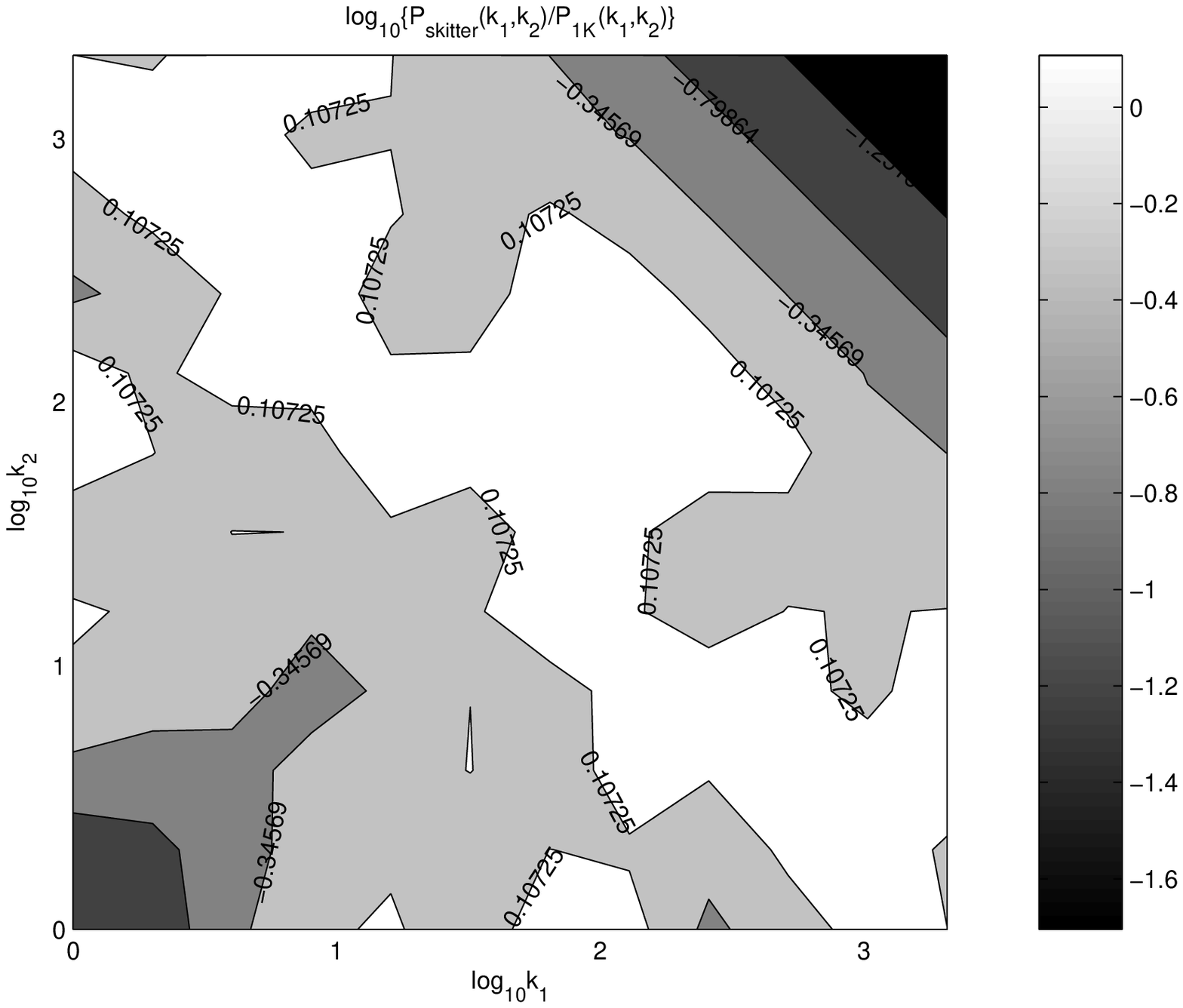}
        \label{fig:pkk-skitter}}
        \hfill
        \subfigure[BGP tables]
        {\includegraphics[width=1.7in]{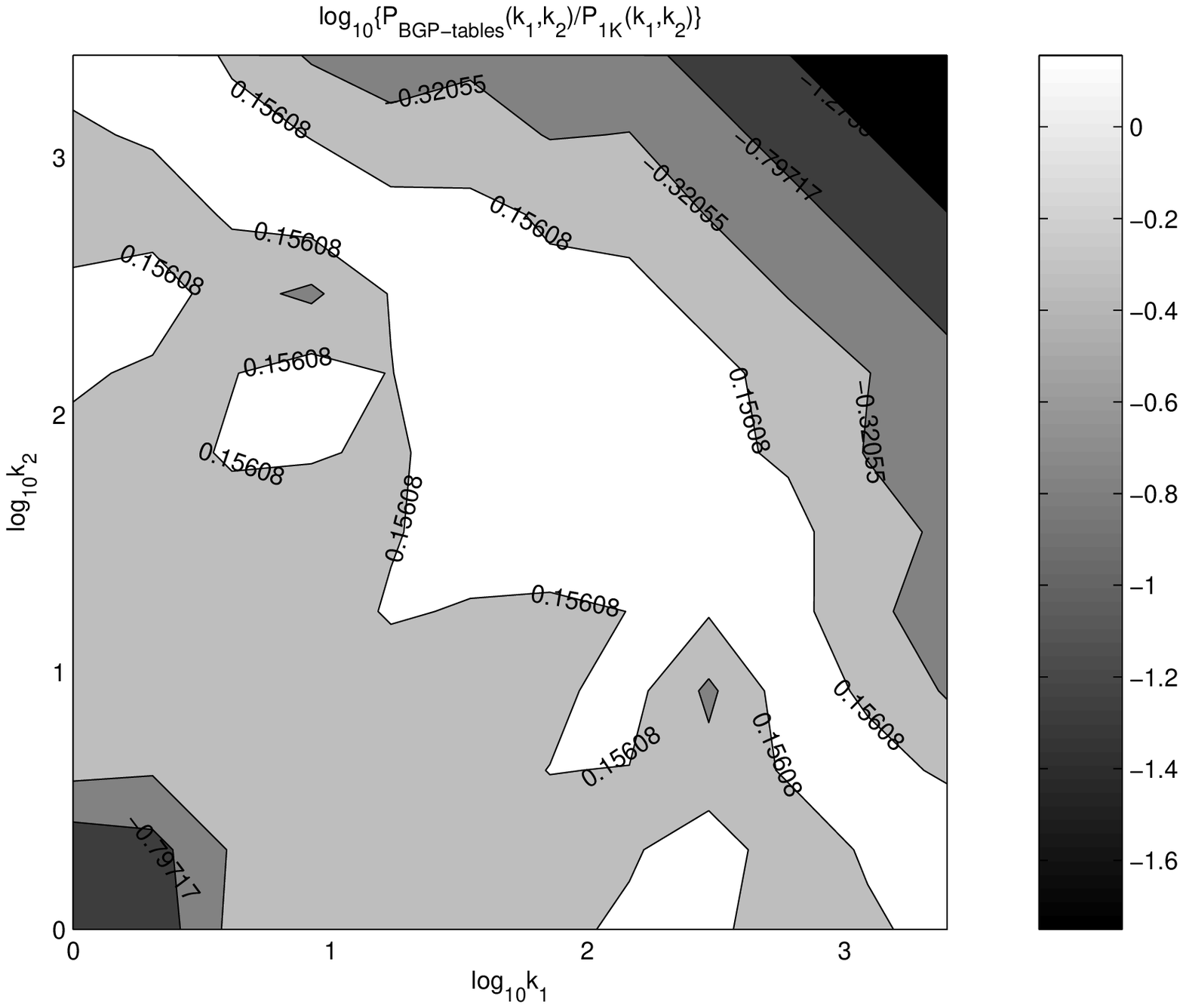}
        \label{fig:pkk-bgp}}
        \hfill
        \subfigure[WHOIS]
        {\includegraphics[width=1.7in]{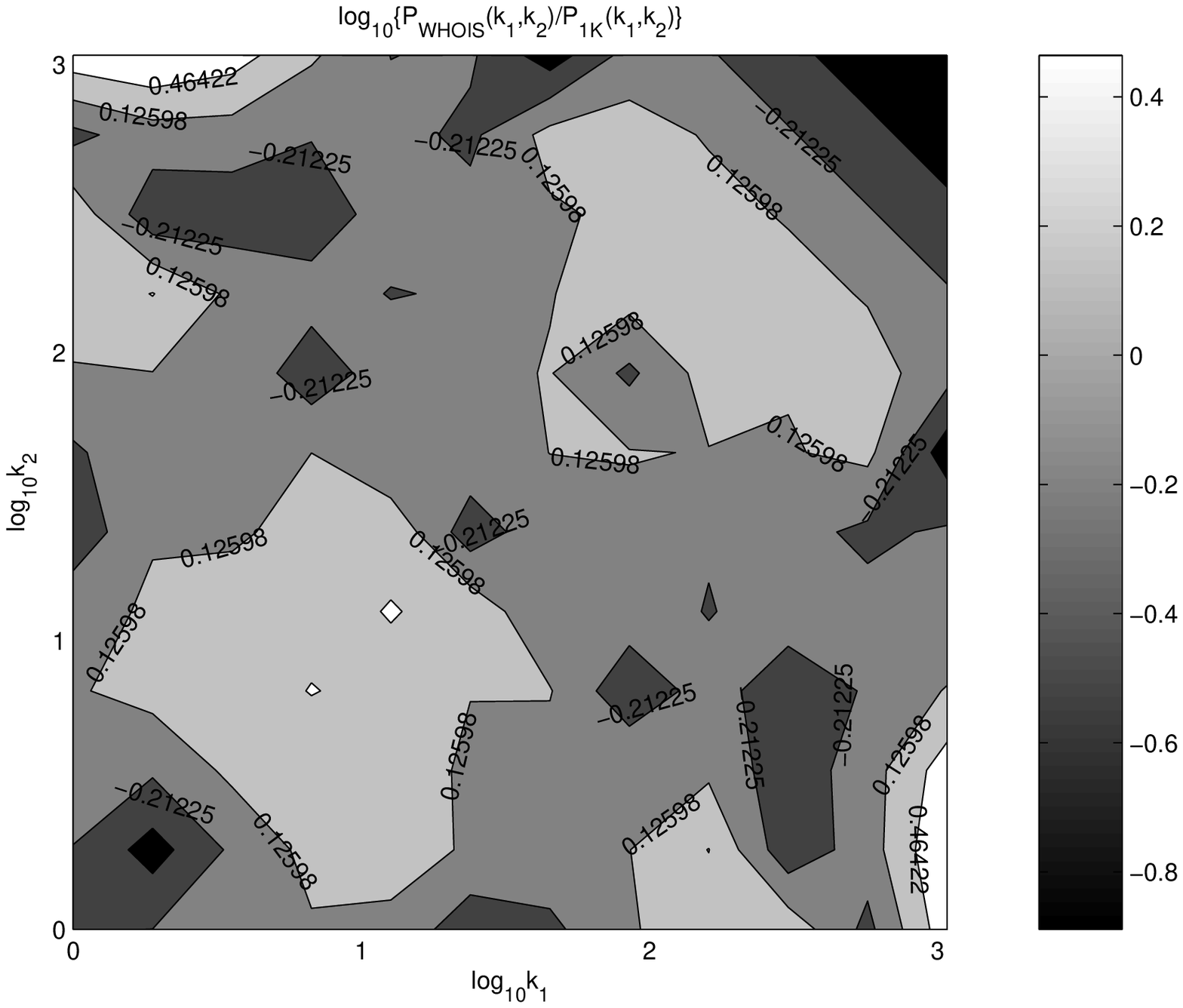}
        \label{fig:pkk-whois}}
    }
    \caption{\footnotesize {\bf Comparison of graphs with 1K-random models.}
    {\bf a)} The contour plot of the logarithm of the joint degree
    distribution~$P_{1K}(k_1,k_2)$ for a 1K-random graph having the skitter
    degree distribution~$P(k)$.
    {\bf b)} The logarithm of the ratio of~$P(k_1,k_2)$ observed in the real
    skitter graph to its simulated~$P_{1K}(k_1,k_2)$.
    {\bf c, d)} The plots, analogous to (b), for BGP and WHOIS graphs.
    Some asymmetry of the diagrams is due to interpolation and rounding algorithms
    in MATLAB. The {\em scatter\/} plots in the Supplement~\cite{comp-anal} are
    symmetric.
}
    \label{fig:pkk}
\end{figure*}

For each of our graphs, we consider its 1K-random counterpart reproducing $P(k)$
of the given graph. We calculate the JDD of the model and compare it with the
actual JDDs of our graphs (Figure~\ref{fig:pkk}).

The 1K-random graph generated from skitter's node degree distribution
(Figure~\ref{fig:ppk-skitter}) has the smallest frequency of tangential links
interconnecting medium-degree nodes (the minimum in the center of the plot).
The most frequent links are either radial (bottom-right and top-left corners) or
low-degree tangential (bottom-left corner). The ratio of the actual JDD of the
skitter graph to this model (Figure~\ref{fig:pkk-skitter}) shows that the real
skitter topology is quite different from its 1K-random version. The actual skitter
graph exhibits a relative deficiency of links in the core and in the fringe
(minimum of the ratio in the top-right and bottom-left corners). At the same
time, it has a relative excess of radial links (bottom-right and top-left
corners) and of tangential links in the medium-degree zone (the center of the plot).

The ratio of the BGP graph JDD to its 1K-random counterpart is similar to skitter
ratio, but the excess of radial links is less prominent (Figure~\ref{fig:pkk-bgp}).
The ratio of the WHOIS graph JDD to its 1K-random model is less variable
(Figure~\ref{fig:pkk-whois}) showing that the WHOIS graph is closer to being
1K-random than the other two graphs.

We now turn our attention to other JDD-derived statistics
(cf.~Section~\ref{sec:joint-degree}). The assortativity coefficient of uncorrelated 1K- and 0K-random graphs is $r=0$
and that their average neighbor connectivity~$k_{nn}(k)$ is a
constant function of node degree~$k$~\cite{DorMen-book03}. For 1K-random graphs, it
is~\mbox{$k_{nn}^{1K}(k)=\langle k^2 \rangle/\bar{k}$},
where~\mbox{$\langle k^2 \rangle$} denotes the second moment of the degree
distribution. For 0K-random graphs, the expression is:~\mbox{$k_{nn}^{0K}(k)=\bar{k}+1$}.
While all three of our data sources yield disassortative graphs
with~\mbox{$r<0$}, the assortativity coefficient of the WHOIS graph is closest
to~$0$ (cf.~Table~\ref{table:summary}). Its average neighbor degree~$k_{nn}(k)$
varies within a factor of 2. In contrast, the average neighbor degree of the
other two graphs varies by two orders of magnitude (cf.~Figure~\ref{fig:knnk}).
These observations again point out that the WHOIS graph is the closest to being
1K-random. Note, however, that PLRG-generated graphs~\cite{AiChLu00,TaGoJaShWi02}
cannot accurately approximate the WHOIS topology since its degree distribution
does not follow power-law.

The skitter graph is on the other extreme: it is the most disassortative (the
smallest value of~$r$) and its average neighbor degree~$k_{nn}(k)$ has the
sharpest decline (the largest value of exponent~$\gamma_{nn}$ of the power-law
fit of~$k_{nn}(k)$). In other words, even though this graph has a power-law
degree distribution, the 1K-random (PLRG) model cannot accurately approximate
it either.

\subsubsection{Clustering}\label{sec:2k-clustering}
In this section, we focus on how clustering can be used to verify the accuracy of topology models.
Uncorrelated graphs have not only constant average neighbor connectivity but
also constant clustering. For 1K-random graphs, it
is:~\mbox{$C_{1K} = (\langle k^2 \rangle - \bar{k}^2)/(n\bar{k}^3)$},
while for 0K-random graphs, we
have~\mbox{$C_{0K} = \bar{k}/n$}~\cite{DorMen-book03}.
Dorogovtsev~\cite{dorogovtsev04} showed that the 2K-random graphs have a
specific form of local clustering~\mbox{$C_{2K}(k)$} and derived expressions
for mean local clustering~\mbox{$\bar{C}_{2K}$} and clustering
coefficient~\mbox{$C_{2K}$} (Eqs.~(8),~(9), and~(10) in~\cite{dorogovtsev04},
correspondingly).

\begin{figure*}[tbh]
    \centerline{
        \subfigure[skitter]
        {\includegraphics[width=2.25in]{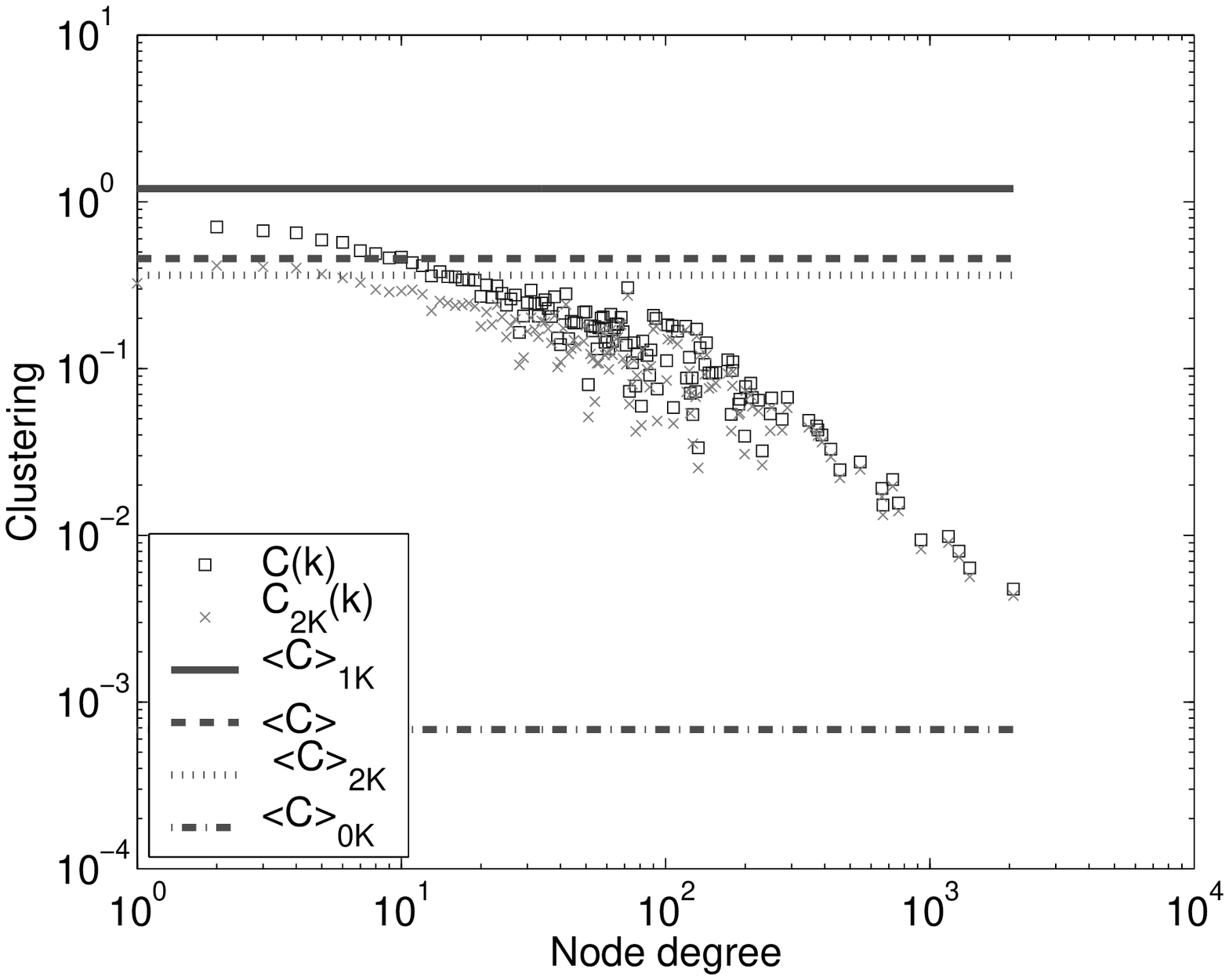}
        \label{fig:clus-skitter}}
        \hfill
        \subfigure[BGP tables]
        {\includegraphics[width=2.25in]{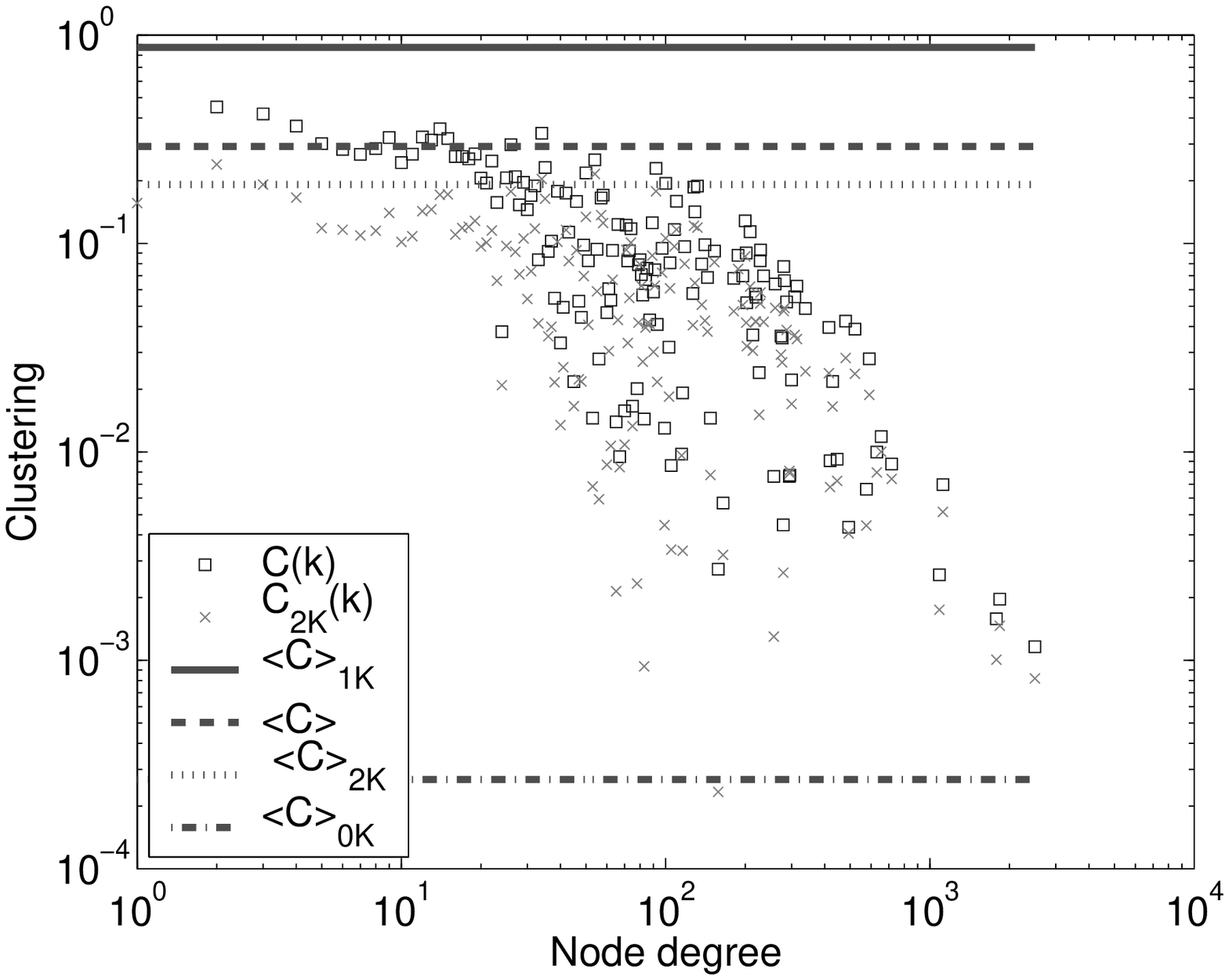}
        \label{fig:clus-bgp}}
        \hfill
        \subfigure[WHOIS]
        {\includegraphics[width=2.25in]{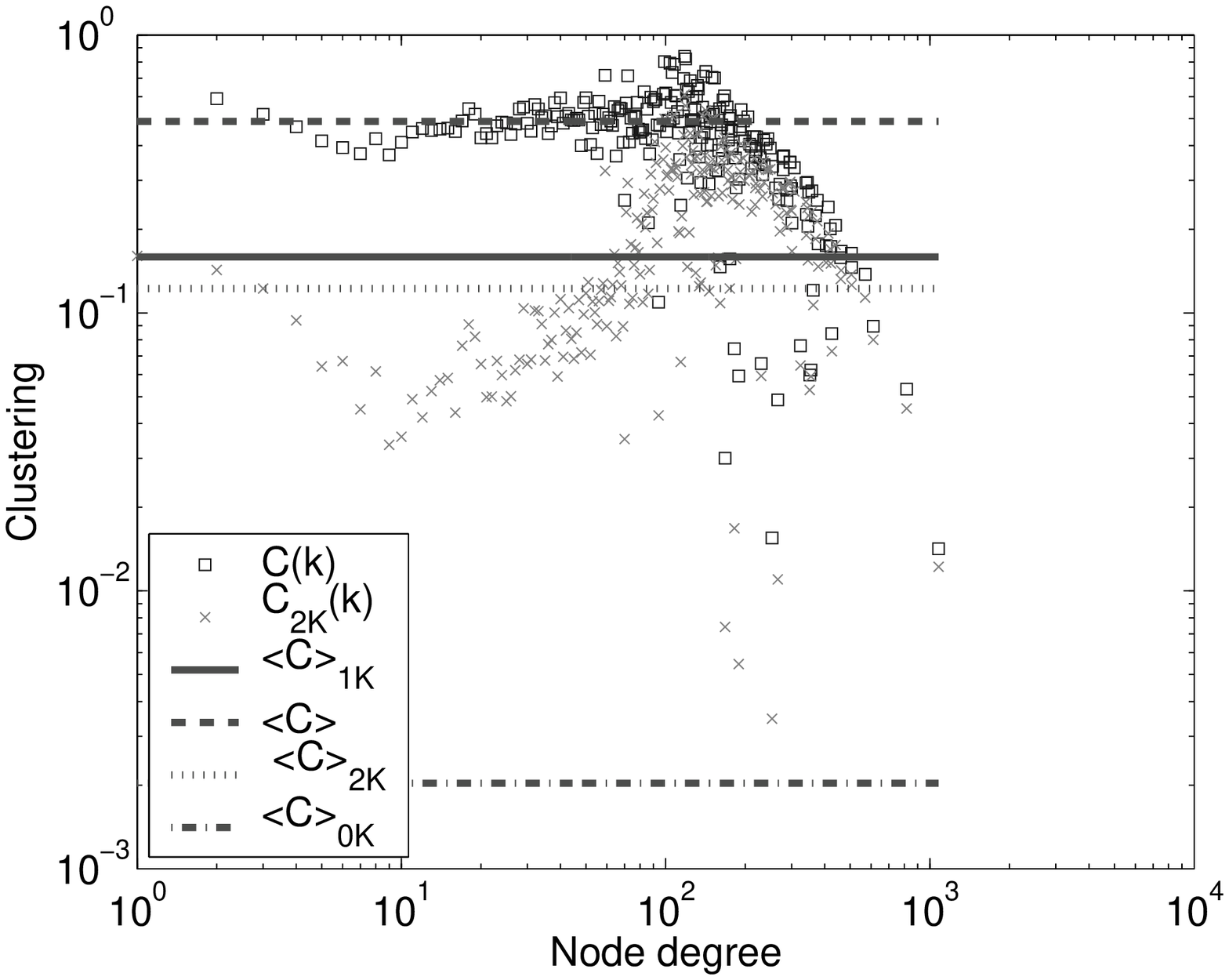}
        \label{fig:clus-whois}}
    }
    \caption{\footnotesize {\bf Local clustering vs.~graph randomness.}
    Squares show local clustering observed in the real
    topology. The dashed line is its mean value.
    Crosses show local clustering predicted by the
    2K-random graph model~\mbox{$C_{2K}(k)$}. The dotted line
    is its mean value. The solid and dash-dotted lines
    are constant clusterings predicted by the 1K- and
    0K-random graph models.
    }
    \label{fig:clustering_vs_randomness}
\end{figure*}

We compare clustering observed in our three Internet graphs with the
predicted values for different graph models (Figure~\ref{fig:clustering_vs_randomness}).
In the skitter and BGP cases, the local clustering function~$C_{2K}(k)$
calculated for the 2K-random model follows, albeit shifted down, the form of actually observed clustering~$C(k)$. The ratio of corresponding
mean values~\mbox{$\bar{C}_{2K}/\bar{C}$} is 0.8 for the skitter graph and
0.7 for the BGP graph. In the WHOIS case, the functional behavior of the
model and of the observed clustering are different, and the ratio of
their mean values is 0.25. We conclude that, using the metric of clustering,
the skitter graph is closest to being 2K-random, while the WHOIS graph is the
furthest. This finding has a direct impact on topology generators: it implies that the
skitter topology can be successfully recreated by capturing the JDD observed in the measured topology. We surmise that a 2K-random generator will closely approximate the skitter graph.
Similarly, a 2K-random generator
reproducing the JDD observed in the measured BGP graph will be able to
create an approximate model of the BGP graph.

Figure~\ref{fig:clustering_vs_randomness} also shows the constant
values of local clustering predicted by the corresponding 1K- and 0K-random
graph models, $C_{1K}$ (solid line) and $C_{0K}$ (dash-dotted line).
Naturally, the 1K-random graphs, with a constant form of local clustering, less
accurately describe the observed clustering than 2K-random model, except in the
WHOIS case, which is closest to being 1K-random.
Clustering in the 0K-random graphs is even further away, being
orders of magnitude smaller than the clustering observed in all three graphs.
Note that the ratio of~\mbox{$\bar{C}_{0K}/\bar{C}_{1K}$} is an indirect
indicator of a graph's proximity to being 0K-random.
The~\mbox{$\bar{C}_{0K}/\bar{C}_{1K}$} values for our graphs ($1 \cdot 10^{-2}$
for the WHOIS, $6 \cdot 10^{-4}$ for the skitter, and $3 \cdot 10^{-4}$ for the
BGP) indicate that the WHOIS graph is better approximated by 0K-random model,
compared with the other two graphs. The BGP graph is the least 0K-random in that
respect.

In summary, the 2K-random graph model approximates the skitter topology best,
while the PLRG generator is inferior for all the three graphs.

\section{Limitations}\label{sec:limitations}

Our work suffers from a number of methodological limitations and
biases.  We discuss each in turn below along with the potential
consequences.

We have tried to be exhaustive while compiling our list of graph
metrics considered by the community.  However, it is possible that we
may have missed important metrics or that additional important metrics
may be proposed that are not well captured by, for instance, joint
degree distribution.

Another limitation is our available data. Although the data sets we
examine represent the current state of the art in macroscopic AS topology,
they are incomplete and indirect reflections of the underlying
topology. They require processing before producing the desired AS
graph. For all our data sets, researchers must make choices
while dealing with ambiguities and errors in the raw data. One such
example is the detection of ``false'' links created by route changes in
traceroute data.
This paper does not address how different choices in processing the original data may result in
different values for our target metrics.  Instead we have attempted,
where possible, to use best practices to extract topologies as presented in papers and in
our discussions with other researchers.

Next, we limit our data collection to a single month for obtaining skitter and BGP data. While we believe that our results will hold true
for historical data and are not an artifact of the current Internet or our sampling period, we leave this study to future work.

Finally, we come to the role played by JDD in topological studies. JDD
has successfully explained the resulting metric values as well as inherent
differences in skitter, BGP and WHOIS graphs. As a next step, we
compare clustering in our observed topologies to the predicted
clustering values in the 2K-random graph. The proximity between the
observed and predicted models gives us further reason to believe that
graphs generated by the 2K-random model come close to the original
topology. Ideally, we could use a graph generator that uses the
measured JDD of a graph to produce random graphs with similar JDDs,
which in turn would also display similar values for a variety of
important graph metrics.  We leave such a potential demonstration of
the value of JDD for capturing a broad range of graph characteristics
to future work.

\section{Analysis and Conclusions}\label{sec:conclusion}

We discussed the properties of Internet AS-level topologies extracted
from the three most popular sources of AS topology data: skitter
measurements, BGP tables, and the RIPE WHOIS database. We compared the
derived topologies based on a set of important and frequently used
statistical characteristics.

We further presented a detailed comparison of widely available sources
of topology data in terms of a number of popular metrics studied in
the literature.  Of the set of metrics we considered, the joint degree
distribution~\mbox{$P(k_1,k_2)$} embeds the most information about a
graph, since this distribution determines both the average node
degree~$\bar{k}$ and the assortativity coefficient~$r$. We find that,
for the data sources we consider, {\em a 2K-random model reproducing
the JDD of the original topology also captures other crucial
topological characteristics.}  While additional work is required to
verify this claim, we believe that JDD may be a powerful metric for
capturing a variety of important graph properties.  Isolating such a
metric or small set of metrics is a prerequisite to developing a
accurate topology generators to assist a broad array of research and
development efforts.  Developing such a JDD-based topology generator
and further demonstrating this concept is the subject of our current
research.

We also propose criteria to evaluate how well the random graph models
reproducing the average node degree~$\bar{k}$ (0K-random), the degree
distribution~\mbox{$P(k)$} (1K-random), or the JDD~\mbox{$P(k_1,k_2)$}
(2K-random) approximate characteristics of the observed
topologies. Using clustering as a measure of accuracy of the 2K-random
approximation, we find that the 2K-random model describes the skitter
graph most accurately. Using the assortativity coefficient (calculated
from the JDD) as a measure of accuracy of the 1K-random approximation,
we find that 1K- or 0K-random graph descriptions best fit the WHOIS
graph, but are less successful in the skitter and BGP cases. The
latter fact implies that the power law random graph (PLRG) model
(which is a special case of 1K-random models) and topology generators
based on it fail to accurately capture the important properties of the
skitter or BGP graphs. Similarly, the PLRG model fails to recreate the
WHOIS graph since its node degree distribution does not follow a power
law at all.

Finally, one may ask which data source is closest to reality. We
emphasize that there is not one but at least {\em three\/} data
sources of the Internet AS-level topology: skitter, BGP, and WHOIS
data, and that the resulting graphs present different views of the
Internet.  The skitter graph closely reflects the topology of actual
Internet traffic flows, i.e.~the data plane. The BGP graph reveals the
topology seen by the routing system, i.e.~the control plane.
Naturally, these two topologies are somewhat different. Understanding
their incongruities is a subject of ongoing
research~\cite{HyuBroCla03,MaReWaKa03,MaJoReWaKa04}. The WHOIS graph
represents a record of the Internet topology created by human actions,
i.e.~the management plane. It is not surprising that this
human-generated view of the Internet has different topological
properties than the other two graphs. The observed abundance of
tangential links between ASes is likely to reflect unintentional or
even intentional over-reporting by some providers of their peering
arrangements.

Our analysis should arm researchers with better insights into specifics of each topology. We hope
that our study encourages the validation of existing models against real data and also motivates t
he development of  better topology models.

\begin{table*}[tbh]
    \centering
    \caption{\bf Summary statistics.
     }
    \begin{tabular}{|c|c|c|c|c|}
\cline{3-5}
 \multicolumn{2}{c|}{ }  & skitter  & BGP tables  & WHOIS \\ \hline
 Average degree & Number of nodes {\footnotesize $(n)$ } & 9,204 & 17,446 & 7,485 \\ \cline{2-5}
  & Number of edges {\footnotesize $(m)$ } & 28,959 & 40,805 & 56,949 \\ \cline{2-5}
  & Avg node degree {\footnotesize  $(\bar{k})$ } & 6.29 & 4.68 & 15.22 \\ \cline{2-5}
\hline
 Degree distr & Max node degree  {\footnotesize $(k_{max})$ } & 2,070 & 2,498 & 1,079 \\ \cline{2-5}
  & Power-law max degree { \footnotesize $(k_{max}^{PL}) $ } & 1,448 & 4,546 & - \\ \cline{2-5}
  & Exponent of {\footnotesize $P(k)\;(-\gamma)$ } & 2.25 & 2.16 & - \\ \cline{2-5}
\hline
 Joint degree distr & Avg neighbor degree {\footnotesize $(\bar{k}_{nn}/(n-1))$ } & 0.05 & 0.03 & 0.02 \\ \cline{2-5}
  & Exponent of {\footnotesize $k_{nn}(k)\;(-\gamma_{nn})$ } & 1.49 & 1.45 & - \\ \cline{2-5}
  & Assortative coefficient {\footnotesize $(r)$ } & -0.24 & -0.19 & -0.04 \\ \cline{2-5}
\hline
 Clustering & Mean clustering {\footnotesize $(\bar{C})$ } & 0.46 & 0.29 & 0.49 \\ \cline{2-5}
  & Clustering coefficient {\footnotesize $(C)$ } & 0.03 & 0.02 & 0.31 \\ \cline{2-5}
  & Exponent of {\footnotesize $C(k)\;(-\gamma_{C})$} & 0.33 & 0.34 & - \\ \cline{2-5}
\hline
Rich club  & Exponent of {\footnotesize $\phi(\rho/n)\;(-\gamma_{rc})$} & 1.48 & 1.45 & 1.69 \\ \cline{2-5}
\hline
 Coreness & Avg node coreness  {\footnotesize $(\bar{\kappa})$ } & 2.23 & 1.41 & 7.65 \\ \cline{2-5}
  & Max node coreness  {\footnotesize $(\kappa_{max})$ } & 27 & 27 & 87 \\ \cline{2-5}
  & Core size ratio {\footnotesize $(n_{core}/n)$ } & $5  \cdot 10^{-3}$ & $3 \cdot 10^{-3}$ & $17 \cdot 10^{-3} $\\ \cline{2-5}
  & Min degree in core {\footnotesize $(k_{core}^{min})$ } & 68 & 34 & 99 \\ \cline{2-5}
  & Fringe size ratio {\footnotesize $(n_{fringe}/n)$ } & 0.27 & 0.29 & 0.06 \\ \cline{2-5}
  & Max degree in fringe {\footnotesize $(k_{fringe}^{max})$} & 5 & 7 & 4 \\ \cline{2-5}
  & Exponent of {\footnotesize $\kappa(k)\;(\gamma_\kappa)$} & 0.68 & 0.58 & 1.07 \\ \cline{2-5}
\hline
 Distance & Avg distance {\footnotesize $(\bar{d})$} & 3.12 & 3.69 & 3.54 \\ \cline{2-5}
  & Std deviation of distance {\footnotesize $(\sigma)$} & 0.63 & 0.87 & 0.80 \\ \cline{2-5}
  & Exponent of {\footnotesize $d(k)\;(-\gamma_d)$} & 0.07 & 0.07 & 0.09 \\ \cline{2-5}
\hline
 Eccentricity & Graph radius {\footnotesize $(R,\;\varepsilon_{min})$} & 4 & 5 & 4 \\ \cline{2-5}
  & Avg eccentricity {\footnotesize $(\bar{\varepsilon})$} & 5.11 & 6.61 & 6.12 \\ \cline{2-5}
  & Graph diameter {\footnotesize $(D,\;\varepsilon_{max})$} & 7 & 10 & 8 \\ \cline{2-5}
  & Center size ratio {\footnotesize $(n_R/n)$} & $320 \cdot 10^{-4}$ & $14  \cdot 10^{-4}$ & $1 \cdot 10^{-4}$ \\ \cline{2-5}
  & Min degree in center {\footnotesize $(k_R^{min})$} & 4 & 188 & 1,079 \\ \cline{2-5}
  & Periphery size ratio {\footnotesize $(n_D/n)$} & $21  \cdot 10^{-4}$& $2 \cdot 10^{-4}$ & $106  \cdot 10^{-4}$ \\ \cline{2-5}
  & Max degree in periphery {\footnotesize $(k_D^{max})$} & 1 & 1 & 6 \\ \cline{2-5}
\hline 
 Betweenness & Avg node betweenness {\footnotesize $(\bar{B}_{node}/(n(n-1)))$} & $11 \cdot 10^{-5}$  & $7.7 \cdot 10^{-5} $ & $17 \cdot 10^{-5}$ \\ \cline{2-5}
  & Exponent of {\footnotesize $B(k)\;(\gamma_B)$ } & 1.35 & 1.17 & - \\ \cline{2-5}
  & Avg edge betweeness  {\footnotesize $(\bar{B}_{edge}/(n(n-1)))$ } & $5.37 \cdot 10^{-5} $ & $4.51 \cdot 10^{-5} $ & $3.10 \cdot 10^{-5} $ \\ \cline{2-5}
\hline
 Spectrum & Largest eigenvalue & 79.53 & 73.06 & 150.86 \\ \cline{2-5}
  & Second largest eigenvalue & -53.32 & -55.13 & 68.63 \\ \cline{2-5}
  & Third largest eigenvalue & 36.40 & 53.54 & 62.03 \\ \cline{2-5}
\hline
\end{tabular}

    \label{table:summary}
\end{table*}

\section{Acknowledgments}

We thank Ulrik Brandes for sharing his betweenness code with us
and Andre Broido for answering our questions.

Support for this work was provided by NSF CNS-0434996,
NCS ANI-0221172, Cisco's University Research program,
and other CAIDA members.

\bibliographystyle{IEEE}
\footnotesize{
\bibliography{bib}

\begin{thebibliography}{10}

\bibitem{FaFaFa99}
M.~Faloutsos, P.~Faloutsos, and C.~Faloutsos,
\newblock ``On power-law relationships of the {Internet} topology,''
\newblock in {\em ACM SIGCOMM}, 1999, pp. 251--262.

\bibitem{WilRevisited}
Q.~Chen, H.~Chang, R.~Govindan, S.~Jamin, S.~J. Shenker, and W.~Willinger,
\newblock ``The origin of power laws in {Internet} topologies revisited,''
\newblock in {\em IEEE INFOCOM}, 2002.

\bibitem{TaGoJaShWi02}
H.~Tangmunarunkit, R.~Govindan, S.~Jamin, S.~Shenker, and W.~Willinger,
\newblock ``Network topology generators: Degree-based vs. structural,''
\newblock in {\em ACM SIGCOMM}, 2002, pp. 147--159.

\bibitem{LiAlWiDo04}
L.~Li, D.~Alderson, W.~Willinger, and J.~Doyle,
\newblock ``A first-principles approach to understanding the {Internet's}
  router-level topology,''
\newblock in {\em ACM SIGCOMM}, 2004.

\bibitem{BuTo02}
T.~Bu and D.~Towsley,
\newblock ``On distinguishing between {Internet} power law topology
  generators,''
\newblock in {\em IEEE INFOCOM}, 2002.

\bibitem{JaRoTo04}
S.~Jaiswal, A.~L. Rosenberg, and D.~Towsley,
\newblock ``Comparing the structure of power-law graphs and the {Internet AS}
  graph,''
\newblock in {\em IEEE ICNP}, 2004.

\bibitem{GaPa04}
M.~Gaertler and M.~Patrignani,
\newblock ``Dynamic analysis of the {Autonomous System} graph,''
\newblock in {\em IPS}, 2004.

\bibitem{ZhoMo04}
S.~Zhou and R.~J. Mondrag\'{o}n,
\newblock ``Accurately modeling the {Internet} topology,''
\newblock {\em Physical Review E}, vol. 70, pp. 066108, 2004,
\newblock \url{http://arxiv.org/abs/cs.NI/0402011}.

\bibitem{DorMen-book03}
S.~N. Dorogovtsev and J.~F.~F. Mendes,
\newblock {\em Evolution of Networks: From Biological Nets to the {Internet}
  and {WWW}},
\newblock Oxford University Press, Oxford, 2003.

\bibitem{nets-nr-won}
{CAIDA},
\newblock ``Toward mathematically rigorous next-generation routing protocols
  for realistic network topologies,'' Research Project,
\newblock \url{http://www.caida.org/projects/nets-nr/}.

\bibitem{vahdat02}
A.~Vahdat, K.~Yocum, K.~Walsh, P.~Mahadevan, D.~Kostic, J.~Chase, and
  D.~Becker,
\newblock ``Scalability and accuracy in a large-scale network emulator,''
\newblock in {\em {OSDI}}, 2002.

\bibitem{skitter}
kc~claffy, T.~E. Monk, and D.~McRobb,
\newblock ``Internet tomography,''
\newblock {\em Nature}, January 1999,
\newblock \url{http://www.caida.org/tools/measurement/skitter/}.

\bibitem{routeviews}
``{University of Oregon RouteViews Project},''
  \url{http://www.routeviews.org/}.

\bibitem{irr}
``{Internet Routing Registries},'' \url{http://www.irr.net/}.

\bibitem{comp-anal}
{CAIDA},
\newblock ``Comparative analysis of the {Internet} {AS}-level topologies
  extracted from different data sources: Data page,''
  \url{http://www.caida.org/analysis/topology/as_topo_comparisons/}.

\bibitem{traceroute}
``{\tt traceroute},'' \url{http://www.traceroute.org/#source%20code}.

\bibitem{as-adjacencies}
{CAIDA},
\newblock ``Macroscopic topology {AS} adjacencies,''
  \url{http://www.caida.org/tools/measurement/skitter/as_adjacencies.xml}.

\bibitem{MaReWaKa03}
Z.~M. Mao, J.~Rexford, J.~Wang, and R.~H. Katz,
\newblock ``Towards an accurate {AS}­-level traceroute tool,''
\newblock in {\em ACM SIGCOMM}, 2003.

\bibitem{bgp}
Y.~Rekhter and T.~Li,
\newblock {\em {A Border Gateway Protocol 4 (BGP-4)}},
\newblock IETF, RFC 1771, 1995.

\bibitem{as-guidelines}
J.~Hawkinson and T.~Bates,
\newblock {\em Guidelines for Creation, Selection, and Registration of an
  {Autonomous System~(AS)}},
\newblock IETF, RFC 1930, 1996.

\bibitem{SiFa04}
G.~Siganos and M.~Faloutsos,
\newblock ``Analyzing {BGP} policies: Methodology and tool,''
\newblock in {\em IEEE INFOCOM}, 2004.

\bibitem{ChaGoJaSheWi04}
H.~Chang, R.~Govindan, S.~Jamin, S.~J. Shenker, and W.~Willinger,
\newblock ``Towards capturing representative {AS}-level {Internet}
  topologies,''
\newblock {\em Computer Networks Journal}, vol. 44, pp. 737--755, April 2004.

\bibitem{skitter-poster}
{CAIDA},
\newblock ``Visualizing {Internet} topology at a macroscopic scale,''
  \url{http://www.caida.org/analysis/topology/as_core_network/}.

\bibitem{LaByCroXie03}
A.~Lakhina, J.~Byers, M.~Crovella, and P.~Xie,
\newblock ``Sampling biases in {IP} topology measurements,''
\newblock in {\em IEEE INFOCOM}, 2003.

\bibitem{DaAlHaBaVaVe05}
L.~Dall'Asta, I.~Alvarez-Hamelin, A.~Barrat, A.~V\'{a}zquez, and A.~Vespignani,
\newblock ``Exploring networks with traceroute-like probes: Theory and
  simulations,''
\newblock {\em Theoretical Computer Science, Special Issue on Complex
  Networks}, 2005,
\newblock \url{http://arxiv.org/abs/cs.NI/0412007}.

\bibitem{as-ranking}
{CAIDA},
\newblock ``Automated {Autonomous System} ({AS}) ranking,'' Research Project,
\newblock \url{http://www.caida.org/analysis/topology/rank_as/}.

\bibitem{newman02}
M.~E.~J. Newman,
\newblock ``Assortative mixing in networks,''
\newblock {\em Physical Review Letters}, vol. 89, no. 20, pp. 208701, 2002.

\bibitem{dorogovtsev03}
S.~N. Dorogovtsev,
\newblock ``Networks with given correlations,''
  \url{http://arxiv.org/abs/cond-mat/0308336v1}.

\bibitem{WiJa02}
J.~Winick and S.~Jamin,
\newblock ``Inet-3.0: {Internet} topology generator,''
\newblock Technical Report UM-CSE-TR-456-02, University of Michigan, 2002.

\bibitem{BreChaGaRaSi01}
Y.~Breitbart, C.-Y. Chan, M.~Garofalakis, R.~Rastogi, and A.~Silberschatz,
\newblock ``Efficiently monitoring bandwidth and latency in {IP} networks,''
\newblock in {\em IEEE INFOCOM}, 2001.

\bibitem{PaLe01}
K.~Park and H.~Lee,
\newblock ``On the effectiveness of route-based packet filtering for
  distributed {DoS} attack prevention in power-law internets,''
\newblock in {\em ACM SIGCOMM}, 2001.

\bibitem{BoRi02}
B.~Bollob\'{a}s and O.~Riordan,
\newblock ``Mathematical results on scale-free random graphs,''
\newblock in {\em Handbook of Graphs and Networks}, Berlin, 2002, Wiley-VCH.

\bibitem{newman03b}
M.~E.~J. Newman,
\newblock ``Properties of highly clustered networks,''
\newblock {\em Physical Review E}, vol. 68, pp. 026121, 2003.

\bibitem{fraigniaud05}
P.~Fraigniaud,
\newblock ``A new perspective on the small-world phenomenon: Greedy routing in
  tree-decomposed graphs,''
\newblock Technical Report LRI-1397, LRI, University Paris-Sud, 2005.

\bibitem{SoVa04}
S.~N. Soffer and A.~V\'{a}zquez,
\newblock ``Clustering coefficient without degree correlations biases,''
  \url{http://arxiv.org/abs/cond-mat/0409686}.

\bibitem{bollobas85}
B.~Bollob\'{a}s,
\newblock {\em Random Graphs},
\newblock Academic Press, New York, 1985.

\bibitem{harary94}
F.~Harary,
\newblock {\em Graph Theory},
\newblock Addison-Wesley, Reading, MA, 1994.

\bibitem{AlDaBaVe04}
I.~Alvarez-Hamelin, L.~Dall'Asta, A.~Barrat, and A.~Vespignani,
\newblock ``$k$-core decomposition: A tool for the visualization of large scale
  networks,'' \url{http://arxiv.org/abs/cs.NI/0504107}.

\bibitem{peleg01}
D.~Peleg,
\newblock {\em Distributed Computing: A Locality-Sensitive Approach},
\newblock SIAM, Philadelphia, PA, 2000.

\bibitem{KrFaYa04}
D.~Krioukov, K.~Fall, and X.~Yang,
\newblock ``Compact routing on {Internet}-like graphs,''
\newblock in {\em IEEE INFOCOM}, 2004.

\bibitem{brandes01}
U.~Brandes,
\newblock ``A faster algorithm for betweenness centrality,''
\newblock {\em Journal of Mathematical Sociology}, vol. 25, no. 2, pp.
  163--177, 2001.

\bibitem{chung97}
F.~K.~R. Chung,
\newblock {\em Spectral Graph Theory}, vol.~92 of {\em Regional Conference
  Series in Mathematics},
\newblock American Mathematical Society, Providence, RI, 1997.

\bibitem{GkaMiSa03}
C.~Gkantsidis, M.~Mihail, and A.~Saberi,
\newblock ``Conductance and congestion in power law graphs,''
\newblock in {\em ACM SIGMETRICS}, 2003.

\bibitem{VuHuEr01}
D.~Vukadinovi\'{c}, P.~Huang, and T.~Erlebach,
\newblock ``A spectral analysis of the {Internet} topology,''
\newblock Technical Report TIK-NR. 118, ETH, 2001.

\bibitem{GkaMiZe03}
C.~Gkantsidis, M.~Mihail, and E.~Zegura,
\newblock ``Spectral analysis of {Internet} topologies,''
\newblock in {\em IEEE INFOCOM}, 2003.

\bibitem{ErRe59}
P.~Erd\H{o}s and A.~R\'{e}nyi,
\newblock ``On random graphs,''
\newblock {\em Publicationes Mathematicae}, vol. 6, pp. 290--297, 1959.

\bibitem{MolRee95}
M.~Molloy and B.~Reed,
\newblock ``A critical point for random graphs with a given degree sequence,''
\newblock {\em Random Structures and Algorithms}, vol. 6, pp. 161--179, 1995.

\bibitem{MolRee98}
M.~Molloy and B.~Reed,
\newblock ``The size of the giant component of a random graph with a given
  degree sequence,''
\newblock {\em Combinatorics, Probability and Computing}, vol. 7, pp. 295--305,
  1998.

\bibitem{AiChLu00}
W.~Aiello, F.~Chung, and L.~Lu,
\newblock ``A random graph model for massive graphs,''
\newblock in {\em Proceedings of the 32$^{nd}$ Annual ACM Symposium on Theory
  of Computing ({STOC})}. 2000, pp. 171--180, ACM Press.

\bibitem{dorogovtsev04}
S.~N. Dorogovtsev,
\newblock ``Clustering of correlated networks,''
\newblock {\em Physical Review E}, vol. 69, pp. 027104, 2004.

\bibitem{HyuBroCla03}
Y.~Hyun, A.~Broido, and kc~claffy,
\newblock ``Traceroute and {BGP} {AS} path incongruities,''
\newblock in {\em {Cooperative Association for Internet Data Analysis}
  ({CAIDA})}, 2003,
\newblock {\tt http://www.caida.org/outreach/papers/2003/ASP/}.

\bibitem{MaJoReWaKa04}
Z.~M. Mao, D.~Johnson, J.~Rexford, J.~Wang, and R.~Katz,
\newblock ``Scalable and accurate identification of {AS}-level forwarding
  paths,''
\newblock in {\em IEEE INFOCOM}, 2004.

\end{thebibliography}
}
\end{document}